\documentclass[useAMS,usenatbib,usegraphicx]{mn2e}

\def\ltsima{$\; \buildrel < \over \sim \;$}
\def\simlt{\lower.5ex\hbox{\ltsima}}
\def\gtsima{$\; \buildrel > \over \sim \;$}
\def\simgt{\lower.5ex\hbox{\gtsima}}
\def\gsimeq
{\hbox{\raise0.5ex\hbox{$>\lower1.06ex\hbox{$\kern-1.07em{\sim}$}$}}}
\def\lsimeq
{\hbox{\raise0.5ex\hbox{$<\lower1.06ex\hbox{$\kern-1.07em{\sim}$}$}}}

\def\xmm{{\it XMM-Newton }}

\def\xmm{{\it XMM-Newton}}
\def\chandra{{\it Chandra}}

\def\swift{{\it Swift}}
\def\nustar{{\it NuSTAR}}

\def\apj{ApJ}

\def\mnras{MNRAS}
\def\aap{A\&A}

\def\apjl{ApJ}
\def\apjs{ApJS}

\def\nat{Nature}

\def\sgras{Sgr~A$^\star$}

\def\xis{XIS}
\def\xis1{XIS1}
\def\xis2{XIS2}
\def\xis3{XIS3}

\title[] 
 {{Fifteen years of \xmm\ and \chandra\ monitoring of \sgras: 
 Evidence for a recent increase in the bright flaring rate}}

 \author[G.\ Ponti et al. ]
 {G.~Ponti$^{1}$\thanks{ponti@mpe.mpg.de}, 
 B.~De~Marco$^{1}$, 
 M.~R.~Morris$^{2}$, 
 A.~Merloni$^{1}$,  
 T.~Mu\~{n}oz-Darias$^{3,4}$,  
 M.~Clavel$^{5}$, 
 \newauthor
 D.~Haggard$^{6}$, 
 S.~Zhang$^{7}$, 
 K.~Nandra$^{1}$, 
 S.~Gillessen$^{1}$,
 K.~Mori$^{7}$
 J.~Neilsen$^{8}$,
 N.~Rea$^{9,10}$,
 \newauthor
 N.~Degenaar$^{11}$, 
 R.~Terrier$^{12}$ and 
 A.~Goldwurm$^{5,12}$ 
\\ \\
   $^1$ Max Planck Institute fur Extraterrestriche Physik, 85748, Garching, Germany\\
   $^2$ Department of Physics \& Astronomy, University of Calisfornia, Los Angeles, CA 90095-1547, USA \\
   $^3$ Instituto de Astrof\'isica de Canarias, E-38205 La Laguna, Tenerife, Spain \\
   $^4$ Departamento de astrof\'isica, Univ. de La Laguna, E-38206 La Laguna, Tenerife, Spain \\
   $^5$ Service d' Astrophysique/IRFU/DSM, CEA Saclay, B\^at. 709, F-91191 Gif-sur-Yvette Cedex, France \\
   $^6$ Department of Physics and Astronomy, Amherst College, Amherst, MA 01002-5000, USA \\
   $^7$ Columbia Astrophysics Laboratory, Columbia University, New York, NY 10027, USA \\
   $^{8}$ MIT Kavli Institute for Astrophysics and Space Research, Cambridge, MA 02139, USA \\ 
   $^{9}$ Anton Pannekoek Institute for Astronomy, University of Amsterdam, Postbus 94249, NL-1090-GE Amsterdam, The Netherlands \\
   $^{10}$ Institute of Space Sciences (ICE, CSICÐIEEC), Carrer de Can Magrans, S/N, 08193, Barcelona, Spain \\
   $^{11}$ Institute of Astronomy, University of Cambridge, Madingley Road, Cambridge CB3 OHA, UK \\
   $^{12}$ AstroParticule et Cosmologie, Universit\'e Paris Diderot, CNRS/IN2P3, CEA/DSM, Observatoire de Paris, Sorbonne Paris Cit\'e, \\ 
   10 rue Alice Domon et L\'eonie Duquet, F-75205 Paris Cedex 13, France \\
}

\pagerange{\pageref{firstpage}--\pageref{lastpage}}

\usepackage{times}

\begin{document}

\label{firstpage}

\maketitle

\begin{abstract}
We present a study of the X-ray flaring activity of \sgras\ during all the 150 \xmm\ and 
\chandra\ observations pointed at the Milky Way center over the last 15 years. 
This includes the latest \xmm\ and \chandra\ campaigns devoted to monitoring the 
closest approach of the very red Br~$\gamma$ emitting object called G2. 
The entire dataset analysed extends from September 1999 through November 2014.
We employed a Bayesian block analysis to investigate any possible variations in the 
characteristics (frequency, energetics, peak intensity, duration) of the flaring events that 
\sgras\ has exhibited since their discovery in 2001. 
We observe that the total bright-or-very bright flare luminosity of \sgras\ increased between 
2013-2014 by a factor of 2-3 ($\sim3.5 \sigma$ significance). 
We also observe an increase ($\sim99.9$~\% significance) from $0.27\pm0.04$ to 
$2.5\pm1.0$~day$^{-1}$ of the bright-or-very bright flaring rate of \sgras, starting in 
late summer 2014, which happens to be about six months after G2's peri-center passage. 
This might indicate that clustering is a general property of bright flares and that it is 
associated with a stationary noise process producing flares not uniformly distributed in time 
(similar to what is observed in other quiescent black holes). 
If so, the variation in flaring properties would be revealed only now because of the 
increased monitoring frequency. Alternatively, this may be the first sign 
of an excess accretion activity induced by the close passage of G2. 
More observations are necessary to distinguish between these two hypotheses. 
\end{abstract}


\section{Introduction} 

\sgras, the radiative counterpart of the supermassive black hole (BH) 
at the center of the Milky Way radiates currently at a very low rate, about nine orders of 
magnitude lower than the Eddington luminosity for its estimated mass 
of M$_{\rm BH}\sim4.4\times10^6$~M$_{\odot}$ (Ghez et al.\ 2008; 
Genzel et al.\ 2010). The first \chandra\ observations of \sgras\ determined the 
quiescent, absorption-corrected, 2--10~keV X-ray luminosity to be 
L$_{\rm 2-10~keV}\sim2\times10^{33}$~erg~s$^{-1}$ 
(Baganoff et al.\ 2003). This emission is constant in flux, spatially extended and possibly 
due to a radiatively inefficient accretion flow (Rees et al. 1982; Wang et al. 2013).  
On top of the very stable quiescent emission, high-amplitude X-ray flaring activity, 
with variations up to factors of a few hundred times 
the quiescent level is commonly observed (Baganoff et al.\ 2001; Goldwurm et al. 2003; 
Porquet et al.\ 2003; 2008; Nowak et al.\ 2012; Neilsen et al.\ 2013; Degenaar et al. 2013; 
Barri\`ere et al.\ 2014; Haggard et al.\ 2015). 
The most sensitive instruments (e.g., \chandra) established that 
X-ray flares occur on average once per day and they last from several minutes 
up to a few hours, reaching peak luminosities of $\sim5\times10^{35}$~erg~s$^{-1}$. 
In particular, all the observed flares to date have an absorbed power-law spectral 
shape (that will hereinafter be used as our baseline model) consistent with 
a spectral index of 2-2.2 (Porquet et al. 2008; Nowak et al. 2012; 
but see also Barri\`ere et al. 2014). 
Soon after the discovery of the first X-ray flares, the infra-red (IR) counterpart of such 
events was revealed (Genzel et al.\ 2003; Ghez et al. 2004). 
Though every X-ray flare has an IR counterpart, IR flares occur more frequently 
($\sim4$ times higher rate) than the X-ray flares. Moreover, the IR emission 
is continuously variable with no constant level of quiescent emission at low 
fluxes (Meyer et al. 2009). 

The origin of \sgras's flares is still not completely understood. The accreting 
material mostly comes from parts of the stellar winds of the stars orbiting 
\sgras\ (Melia 1992; Coker \& Melia 1997; Rockefeller et al. 2004; 
Cuadra et al.\ 2005; 2006; 2008). The sudden flares might 
be a product of magnetic reconnection, or stochastic acceleration or 
shocks (possibly associated with jets) at a few gravitational radii from \sgras\ 
(Markoff et al. 2001; Liu \& Melia 2002; Liu et al. 2004; Yuan et al. 2003; 2004; 
2009; Marrone et al. 2008; Dodds-Eden et al. 2009). 
Other mechanisms, for instance associated with the tidal disruption of asteroids, 
have also been proposed ({\v C}ade{\v z} et al. 2008; Kosti{\'c} et al. 2009; 
Zubovas et al.\ 2012). To shed light on the radiative mechanism of the flares, 
several multi-wavelength campaigns have been performed (Eckart et al. 2004; 
2006; 2008; 2009; 2012
Yusef-Zadeh et al. 2006; 2008; 2009; Hornstein et al. 2007; Marrone et al. 2008; 
Dodds-Eden et al. 2009; Trap et al.\ 2011). \sgras's spectral energy distribution, 
during flares, shows two peaks of emission, one at IR and the second at X-ray wavelengths. 
The IR peak is consistent with being produced by synchrotron emission (polarisation 
is observed in the submm and IR), while a variety of processes could produce the X-ray 
peak, including synchrotron and inverse Compton processes like synchrotron 
self-Compton and external Compton (see Genzel et al.\ 2010 for a review). 
Synchrotron emission, extending with a break from IR to the X-ray range, 
seems now the best process able to account for the X-ray data with 
reasonable physical parameters (Dodds-Eden et al. 2009, Trap et al. 2010, 
Barri\`ere et al. 2014).

A detailed analysis of the X-ray flare distribution (taking advantage of the 
3~Ms \chandra\ monitoring campaign performed in 2012) shows that 
weak flares are the most frequent, with an underlying power-law flare 
luminosity distribution $dN/dL$ of index $\Gamma\sim-1.9$ (Neilsen et al.\ 2013; 2015). 
In particular, flares with $L_{\rm 2-8~keV}>10^{34}$~erg~s$^{-1}$ occur at a rate of 
$1.1^{+0.2}_{-0.1}$ per day, while luminous flares (with 
$L_{\rm 2-8~keV}>10^{35}$~erg~s$^{-1}$) occur every $\sim10$ days 
(Neilsen et al.\ 2013; 2015; Degenaar et al. 2013). 
The occurrence of flares appears to be randomly distributed and stationary. 
Based on the detection of a bright flare plus three weaker ones during a $\sim230$~ks 
\xmm\ monitoring of \sgras, B\`elanger et al. (2005) and Porquet et al.\ (2008) 
argue that \sgras's flares might occur primarily in clusters. 
An even higher flaring rate was actually recorded during a $\sim23$~ks \chandra\ 
observation (obsID: {\sc 13854}) when 4 weak flares were detected (Neilsen et al.\ 2013; 
with an associated chance probability of about 3.5~\%; Neilsen et al.\ 2015). 
However, no {\it significant} variation of the flaring rate has yet been established. 

Recently, long \chandra, \xmm\ and \swift\ X-ray observing campaigns have 
been performed to investigate any potential variation in \sgras's X-ray properties 
induced by the interaction between \sgras\ and the gas-and-dust-enshrouded G2 object 
(Gillessen et al. 2012; Witzel et al. 2014). We analyse here all the existing \xmm\ 
and \chandra\ observations of \sgras\ to search for variations in the 
X-ray flaring rate. The \swift\ results are discussed elsewhere (Degenaar et 
al. 2015). 

This paper is structured as follows. In \S~2 we summarise the \xmm\ and \chandra\ 
data reduction. In \S~3 we present the \xmm\ monitoring campaigns performed 
in 2013 and 2014. In \S~4 we describe the application of the Bayesian 
block analysis to the 15 years of \xmm\ and \chandra\ data and derive the 
parameters and fluence for each detected flare. We also present the flare fluence 
distribution. In \S~5 we investigate possible variations to the flaring rate of 
\sgras\ and in \S~6 the change in the total luminosity emitted in bright flares. 
Sections 7 and 8 present the discussion and conclusions. 
 
\section{Data reduction} 
\label{datared} 

\subsection{\xmm}

As of November 11, 2014 the \xmm\ archive contains 37 public observations 
that can be used for our analysis of \sgras\footnote{We exclude the observations 
that do not have any EPIC-pn exposures (obsID: 0402430601, 0402430501, 0112971601 
0112972001 and 0505670201), those for which \sgras\ is located 
close to the border of the field of view (obsID: 0112970501 and 0694640401) and 
the observation in timing mode (obsID: 0506291201).}. 
In addition we consider 4 new observations aimed at monitoring the interaction 
between the G2 object and \sgras, performed in fall 2014 (see Tab. \ref{ExpXMM}). 
A total of 41 \xmm\ datasets are considered in this work.  
We reduced the data starting from the {\sc odf} files, using version 13.5.0 
of the \xmm\ {\sc sas} software. 

Several transient X-ray sources are located within a few arcseconds of \sgras, 
contaminating the emission within the corresponding extraction region (of 
$10$~arcsec radius) when they are in outburst. 
There are two such cases in our dataset\footnote{We checked that no flare is 
due to short bursts, such as the type I X-ray bursts from accreting neutron stars 
X-ray binaries, e.g. AX~J1745.6-2901 located at less than 1.5~arcmin from 
\sgras\ (Ponti et al. 2015).}. First, 
CXOGC~J174540.0-290031, an eclipsing low-mass X-ray binary located 
$\sim2.9$~arcsec from \sgras, was discovered by \chandra\ in July 2004 
(Muno et al.\ 2005). This source reached a flux of 
$F_{\rm 2-8~keV}\sim6\times10^{-12}$~erg~cm$^{-2}$~s$^{-1}$ 
(L$_{\rm 2-8~keV}\sim5\times10^{34}$ erg s$^{-1}$) while in outburst, 
significantly contaminating the emission of \sgras\ during 
the \xmm\ observations accumulated in fall 2004 (obsID: 0202670501, 
0202670601, 0202670701 and 0202670801; B\'elanger et al.\ 2005; Porquet et al.\ 2005; 
see Fig. \ref{Slide1}). However this transient contributed no more 
than $\sim50$~\% to the total emission from the \sgras\ extraction region, 
so it did not prevent the detection of bright flares. 
Second, SGR~J1745-2900, the magnetar located $\sim2.4$~arcsec from \sgras\ 
that underwent an X-ray burst on April 25, 2013 (Degenaar et al. 2013; 
Mori et al.\ 2013; Rea et al.\ 2013). 
SGR~J1745-2900 reached a peak flux, just after the outburst, of 
$F_{\rm 1-10~keV}\sim2\times10^{-11}$~erg~cm$^{-2}$~s$^{-1}$, therefore 
dominating the X-ray emission from \sgras's extraction region and preventing 
a clear characterisation of even the brightest flares. Therefore we exclude 
these three observations (obsID: 0724210201, 0700980101 and 0724210501) 
in our present analysis. On the other hand, during the \xmm\ observations 
in fall 2014, the X-ray flux of SGR~J1745-2900 dropped to 
$F_{\rm 1-10~keV}\sim3\times10^{-12}$~erg~cm$^{-2}$~s$^{-1}$ (see Coti Zelati 
et al. 2015; for the details of the decay curve), allowing an adequate characterisation 
of the bright flares (see \S 2.4 and Tab. 2 for the definition of bright flare). 

Due to its higher effective area, this study presents the results obtained 
with the EPIC-pn camera only. We use the EPIC-MOS data (analysed in the 
same way as the EPIC-pn data) to check for consistency. 
Following previous work, we extract the source photons from a circular region 
with $10$~arcsec radius, corresponding to $\sim5.1\times10^4$~AU, or 
$\sim1.3\times10^6$~r$_g$ (r$_g=GM_{\rm BH}/c^2$ being the BH gravitational 
radius, where G is the gravitational constant, $M_{\rm BH}$ is the BH mass and c 
the speed of light; Goldwurm et al. 2003; B\'elanger et al.\ 2005; Porquet et al.\ 2008; 
Trap et al.\ 2011; Mossoux et al.\ 2015). 

Background photons are extracted from a circular region with a radius of $3.5$~arcmin 
located far from the bright diffuse emission surrounding \sgras\ (Ponti et al.\ 2010; 2015a,b). 
Therefore, we typically chose the background regions close to the edge of the field of view. 
Many \xmm\ observations are affected by a high level of particle background activity. 
Despite the small size of the source extraction region, the most intense particle flares 
have a strong effect on the final source light curve, if not filtered out. 
We note that the most intense periods of particle activity occur more 
often towards the start or the end of an orbit, therefore at the start or the end of an 
exposure. To minimize the number of gaps in the final light curve of \sgras\ 
as well as the effect of background variations, we removed the most intense 
period of particle activity, cutting the initial or final part of the exposure, when 
contaminated by bright background flares (see Tab.\ \ref{ExpXMM}). 
We then filtered out the residual flares occurring in the middle of the observation, 
cutting intervals\footnote{The light curves used for this have 20~s bins.} during 
which the 0.3-15~keV light curve exceeded a threshold. 
To decide on a threshold level, we first estimate the fluctuations of the particle 
flares intensity within the detector. To check this, we extracted background 
light curves at different positions from several circular regions with $1'$ radius.
The region positions were chosen to avoid bright sources 
and regions with strong diffuse emission. During the observations 
affected by intense periods of particle activity (such as obsID: 0202670701) 
we observe fluctuations (spatial non-uniformities) by a factor 2-3 between the 
intensities of the background flares observed in the different regions. 
Therefore, a background count rate of about 20~ph~s$^{-1}$ will induce 
$\sim0.04$~ph~s$^{-1}$ in a $10$~arcsec radius circle (the surface ratio between the 
source and background area is 441) and fluctuations of the same order of magnitude. 
Such a value is several times lower than the emission coming from a $10$~arcsec radius 
centred on \sgras\ ($\sim0.2$~ph~s$^{-1}$; quiescent level without spurious sources), 
which guarantees that the final source light curve is not strongly affected by background 
fluctuations. We applied this threshold to all observations. We performed the data 
filtering before running the Bayesian block analysis, to avoid possible biasses 
in our choice of the threshold, to include specific flares. 
{\it A posteriori}, we note that of all bright flares reported in literature, only two 
bright events occurring at the end and beginning of obsID: 0202670601 and 
0202670701, respectively, have been cut (B\'elanger et al. 2005). 

We compute the source and background light curves selecting 
photons in the 2--10~keV band only. Moreover, we selected only single and 
double events using (FLAG == 0) and (\#XMMEA\_EP). 
Source and background light curves have been created using 300~s time bins 
and corrected with the {\sc sas} task {\sc epicclcorr}.
The total EPIC-pn cleaned [and total] exposure corresponds to $\sim1.6$~Ms 
[2.0~Ms]. 

\subsection{\chandra}

We consider here all publicly available \chandra\ observations pointed at \sgras. 
Because of the degradation of the point spread function with off-axis 
angle, we do not consider observations aimed at other sources located 
at a distance more than $2$~arcmin from \sgras. 
All the 46 \chandra\ observations accumulated between 1999 and 2011 and analysed 
here are obtained with the ACIS-I camera without any gratings on (see Tab. \ref{ExpC1}). 
From 2012 onward, data from the ACIS-S camera were also employed.  The 2012 
\chandra\ ''X-ray Visionary Project'' (XVP) is composed of 38 HETG observations with the 
ACIS-S camera at the focus (Nowak et al.\ 2012; Wang et al. 2013; 
Neilsen et al.\ 2013; 2015; see Tab. \ref{ExpC2}\footnote{More information is 
available at this location: www.sgra-star.com}). 
The first two observations of the 2013 monitoring campaign 
were performed with the ACIS-I instrument, while the ACIS-S camera 
was employed in all the remaining observations, after the outburst of 
SGR~J1745-2900 on April 25, 2013. Three observations between May and July 
2013 were performed with the HETG on, while all the remaining ones 
do not employ any gratings\footnote{The ACIS-S instrument, in these last observations,
was used with a subarray mode. In fact, to minimise the CCD frame time, therefore 
reducing the pile-up effect, only the central CCD (S3) with a subarray employing only 128 
rows (1/8 subarray; starting from row number 448) was used (see Tab.\ 
\ref{ExpC}). This resulted in a frame time of 0.4~s for these latter observations.}
(see Tab. 5). 

All the data have been reduced with standard tools from the {\sc ciao} 
analysis suite, version 4.6. Following Neilsen et al. (2013) and Nowak et al. 
(2012), we compute light curves in the 2--8~keV band\footnote{The flare 
fluences, reported in Tab. 3, 4, 5 and 6, are integrated over the 2-10~keV band. } 
and with 300~s time bins. 
Photons from \sgras\ are extracted from a circular region of $1.25$~arcsec radius 
(corresponding to $\sim6400$~AU and $\sim1.6\times10^5$~$r_g$). 
We search for periods of high background levels by creating a light curve 
(of 30~s time bins) from a region of 0.5~arcmin radius, away from \sgras\ and 
bright sources. Periods of enhanced activity are filtered out. 
Thanks to the superior \chandra\ point spread function, less than 
$\sim3$~\% of the flux from SGR~J1745-2900 contaminates the 
extraction region of \sgras, however this is enough to significantly contaminate 
($\sim40$~\%) its quiescent level at the outburst peak. 
We do not correct for this excess flux, however, we note that no flaring activity, 
such as to the one observed in \sgras, is detected in the \chandra\ light curves 
of SGR~J1745-2900. 

\subsubsection{Correction for pile-up}
\label{Spileup}

\begin{figure}
\includegraphics[height=0.47\textwidth,angle=-90]{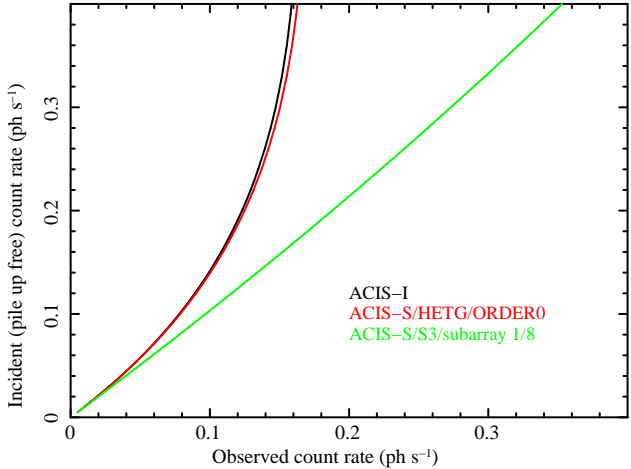}
\caption{Relation between the incident $2-8$~keV count rate (as observed in 
absence of pile-up) and the actual observed $2-8$~keV count rate for three 
different observing modes often used during the \chandra\ observations. 
These relations are derived from the same pile-up models employed in 
the \chandra\ webpimms tool. For the computation we assumed 
the ''absorbed power-law'' spectral model (Nowak et al. 2012) and 
5 active chips (for the ACIS-I and ACIS-S 0$^{th}$-order observations). 
Because of the pile-up effect, with either ACIS-I or ACIS-S with no subarray, 
the observed count rate can not effectively be higher than $0.18$~ph~s$^{-1}$, 
even for incident count rates $>0.2$~ph~s$^{-1}$.}
\label{pileup}
\end{figure}

During the brightest flares the \chandra\ light curves are significantly affected by pile-up, 
if no subarray is used. Figure \ref{pileup} shows the relations between the incident 
count rate, as observed if no pile-up effect is present, and the observed (piled-up) 
count rate. 
This conversion is accurate for the absorbed power-law model. 
We note that the pile-up effect becomes important ($\sim5$~\%) 
for count rates higher than 0.04~ph~s$^{-1}$, well above the quiescent level. 
Therefore, pile-up does not affect the detection of flares or the determination of the 
flaring rate. It does, however, significantly affect the observed peak count rates 
and, therefore, the observed fluences of moderate, bright and especially very 
bright flares. 
Indeed, during either ACIS-I or ACIS-S 0$^{th}$-order observations it is very hard to 
characterise exceptionally bright flares with \chandra\ if no subarray or gratings is 
employed. For example, flares with 
incident peak count rates between $0.25$--$1.0$~ph~s$^{-1}$ would produce an 
observed (piled-up) count rate between $\sim0.14$--$0.17$~ph~s$^{-1}$, never 
higher than $0.18$~ph~s$^{-1}$, if no subarray (or grating) is used (see black and 
red lines in Fig. \ref{pileup}). In particular, we note that the two exceptionally bright flares 
detected in fall 2013 and fall 2014, with peak count rates of $\sim0.5$--$1$~ph~s$^{-1}$, 
respectively, would be heavily piled-up if no subarray were used, giving an observed 
(piled-up) count rate of $\sim0.16$--$0.18$ ~ph~s$^{-1}$. 
Some of the bright flares detected by \chandra, in observations with no subarray, 
could therefore actually be associated with very bright flares. 
The relation shown in Fig. \ref{pileup} is based on the same pile-up model also 
employed in the webpimms\footnote{https://heasarc.gsfc.nasa.gov/cgi-bin/Tools/w3pimms/w3pimms.pl and http://cxc.harvard.edu/toolkit/pimms.jsp} 
tool. We correct the light curves, 
the Bayesian block results (see \S~\ref{SBay}) and the flare fluences 
for the pile-up effect by converting the observed count rates to 
the intrinsic count rates, using the curves shown in Fig. \ref{pileup} 
(see also Tab. \ref{Tabpup}). As it can be seen in Fig. \ref{pileup}, 
as long as the observed count rates are lower than $\sim0.12$~ph~s$^{-1}$ 
the correction for pile-up is accurate, in fact the relations between 
incident and observed count rates are well behaved. For higher count rates 
in observations with no subarray, the relation becomes very steep, 
therefore it becomes increasingly difficult to determine the true incident 
count rate from the observed one. We {\it a posteriori} observe that 
a block count rate higher than 0.12~ph~s$^{-1}$ is observed only during the 
very bright flare observed during obsID 1561 (Baganoff et al. 2001). 

In grating observations, the comparison between the un-piled first-order photons, 
with the  0$^{th}$-order photons provides a recipe to correct count rates and fluences 
for the effect of pile-up, also for the luminous events (see Nielsen et al.\ 2013). 
We observe {\it a posteriori} that our method provide similar results
to the one employed by Nielsen et al. (2013). 

\subsection{Comparison of count rates and fluences between different instruments}

To enable the comparison between the light curves or fluences of flares observed by 
different instruments, we convert the observed corrected count rates and fluences 
from photon numbers into photon energies (in ergs). 
We assume, as established by previous analyses, that all flares have the same 
absorbed power-law shape with spectral index $\Gamma=2$ and are absorbed 
by a column density $N_H=1.5\times10^{23}$~cm$^{-2}$ of neutral material
(Porquet et al. 2008; Nowak et al. 2012). 
Using this model, we convert for each count rate and flaring block (see \S~\ref{SBay}) 
the ''corrected'' count rate and block count rate into a flux (using webpimms$^5$) 
and then use these to compute the fluences in ergs and plot the combined 
light curves (see Fig. 3 and 4). All count rates, fluxes and fluences 
correspond to the absorbed values. 
The value displayed in the last column of Tab. \ref{Tabpup} shows the 
conversion factor. 
\begin{table*}
\small
\begin{center}
\begin{tabular}{ | c c c c c c c }
\hline
\multicolumn{6}{c}{\bf Pile-up correction and conversion factors}\\
Data Mode & p1  & p2 & p3 & p4 & CF \\
\hline
ACIS-I & 1.563 & 1.099 & 1185 & 4.866 & $4.2\times10^{-11}$ \\
ACIS-S HETG 0th & 802.0 & 4.743 & 1.599 & 1.110 & $1.0\times10^{-10}$ \\
ACIS-S HETG 0+1st &  &  &  &  & $5.83\times10^{-11}$\dag \\
ACIS-S 1/8 subarray & 0.2366 & 6.936 & 1.393 & 1.179 & $4.09\times10^{-11}$ \\
EPIC-pn & 1 & 1 & 0 & 0 & $1.3\times10^{-11}$\\ 
\hline
\end{tabular}
\caption{Best fit conversion factors, for each instrument and observing mode, 
between observed count rates (affected by the pile-up effect) into unpiled-up count rates. 
The factors are derived using the webpimms estimates and are in the form: 
$cr_{int}(t)=p1\times cr_o(t)^{p2}+p3\times cr_o(t)^{p4}$, where $cr_o(t)$ is the observed 
count rate and $cr_{int}(t)$ is the intrinsic count rate, once corrected for the pile up effect.
The last column shows the conversion factor (CF) used to transform the corrected count 
rate into a 2--10~keV flux (from ph~s$^{-1}$ into erg~cm$^{-2}$~s$^{-1}$). 
The conversion factor applied to the \chandra\ data is appropriate for unpiled up light 
curves in the 2--8~keV band. \dag See Nowak et al. (2012) and Neilsen et al. (2013). }
\label{Tabpup}
\end{center}
\end{table*} 

\subsection{Classification of flares}

\begin{table}
\small
\begin{center}
\begin{tabular}{ | l c }
\hline
\multicolumn{2}{c}{\bf Definition flares types} \\
{\sc Flare type}          & {\sc Fluence} \\
                                 & ($10^{-9}$~erg~cm$^{-2}$) \\
\hline
{\sc Very bright} & $F>20$ \\
{\sc Bright}         & $5< F\leq 20$ \\
{\sc Moderate}   & $1.5< F\leq 5$ \\
{\sc Weak}         & $F\leq 1.5$ \\ 
\hline
\end{tabular}
\caption{Classification of different flares of \sgras, according to the total 
observed (absorbed) flare fluence in the 2-10~keV energy band. 
We detect 20, 36, 16 and 8 weak, moderate, bright and very bright flares, 
respectively (the number of weak and moderate flares is incomplete). }
\label{DefFlare}
\end{center}
\end{table} 
This work aims at studying the long term trend in \sgras's flaring rate. 
To this end, we consider data from both the \xmm\ and \chandra\ 
monitoring campaigns, regardless of the instrument mode used. 
This has the advantage of increasing the total exposure, therefore 
to provide a larger number of flares. However it has the disadvantage of 
producing an inhomogeneous sample. In fact,  
due to the different background levels of the various cameras and observation 
configurations employed, as well as the diverse point spread functions of the 
different satellites, the detection threshold to weak flares varies between 
observations and, in particular, between different satellites.  
Therefore, we divide the observed flares into four categories (ranked 
according to increasing fluence), {\it weak}, {\it moderate}, {\it bright} and 
{\it very bright} flares.  
The thresholds between the various categories are chosen primarily 
to select homogeneous samples of flares (e.g. observable by all satellites, 
by all instrumental mode of one satellite, etc.), but also to sample the fluence 
distribution with similar portions\footnote{We a-posteriori checked that the 
results presented here do not depend on the details of the choice of these thresholds.}. 
Bright and very bright flares shall be the flares with fluence in excess of 
$5\times10^{-9}$ and $20\times10^{-9}$~erg~cm$^{-2}$, respectively. 
These flares are detectable by both \xmm\ and \chandra, in any observation 
mode employed, given the observed distribution of flare duration (Neilsen et al. 2013). 
Moderate flares are defined as those with fluences between $1.5\times10^{-9}$ 
and $5\times10^{-9}$~erg~cm$^{-2}$. These are easily detectable 
with \chandra\ in any instrument set-up, while the high contribution 
from diffuse emission hampers the detection of a significant fraction of moderate 
flares by \xmm. Therefore, we will only use \chandra\ for their study. 
We consider a weak flare as any significant variation, compared to quiescence, 
with a total fluence lower than $1.5\times10^{-9}$~erg~cm$^{-2}$. 
We note that the various \chandra\ instrumental set ups also have different thresholds 
for the detection of weak flares, with different levels of completeness. 

In summary, observations performed both by \xmm\ and/or \chandra\ 
give us a complete census of bright and very bright flares. 
On the other hand, to have a complete census of moderate flares, we 
restrict ourselves to \chandra\ observations only. 

\section{The 2013-2014 \xmm\ monitoring of SGR~A$^\star$}
\label{envelope}
\begin{figure*}
\includegraphics[height=1\textwidth,width=0.8\textwidth,angle=-90]{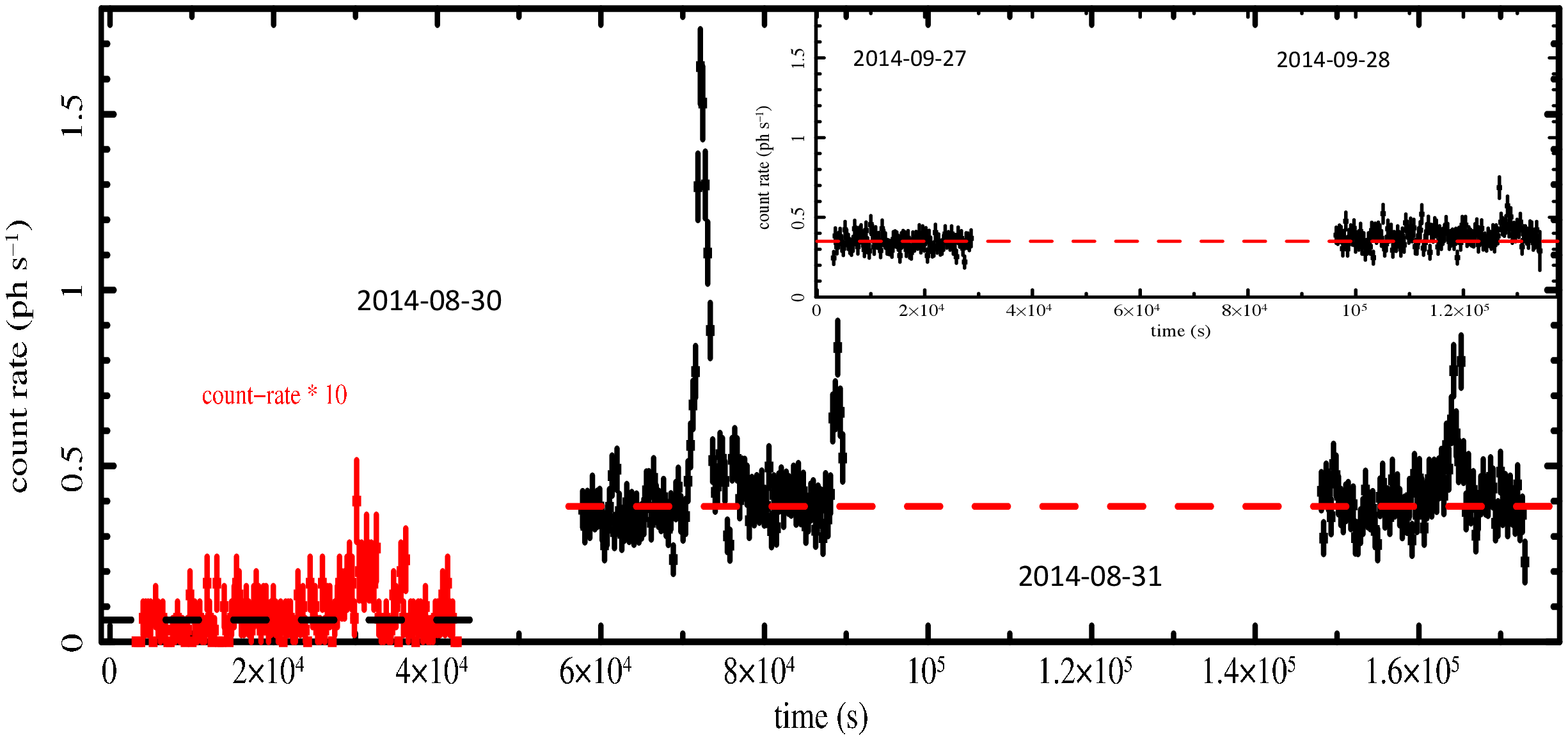}
\vskip-5cm
\caption{\xmm\ and \chandra\ observations accumulated in fall 2014 
are shown in black and red, respectively. For display purposes the 
\chandra\ count rate has been multiplied by 10. Four bright or very bright 
flares are clearly detected within the 4 \xmm\ observations. 
An additional, but weaker, flare is observed during the earlier \chandra\ 
observation. The dashed line shows the best fit of a constant 
to the light curves after excluding the detected flares. 
The constant level observed by \xmm\ follows the decay of the emission 
from SGR~J1745-2900. }
\label{XMM}
\end{figure*}
We start the investigation from the presentation of the new \xmm\ data 
of the intensified monitoring campaign of \sgras, obtained for the peri-center 
passage of G2. 
Three \xmm\ observations were accumulated in fall 2013, in particular, on August 
30, and September 10 and 22. Each light curve can be fitted with a constant 
flux of 0.924, 0.824 and 0.815~ph~s$^{-1}$, respectively. 
We observe no obvious flare activity in any of the three light curves. 
However, our ability to detect moderate or bright flares is hampered by 
the increased flux induced by the outburst of the magnetar SGR~J1745-2900, 
lying within the extraction region of the source light-curve  (see \S~2.1). 
The flux evolution between the different observations follows the typical 
exponential decrease observed in magnetars' outbursts (Rea \& Esposito 2011; 
Coti Zelati et al.\ 2015). Because the dominant contribution from this source 
prevents us from detecting even bright flares from \sgras, we decided 
to discard these observations. 

The light curves of the four \xmm\ observations obtained on August 30, 
31, September 27 and 28, 2014 (black data-points in Fig.\ \ref{XMM}) 
instead show four bright flares (with fluence of 289.4, 62.5, 102.2 and 
$58.8\times10^{-10}$~erg~cm$^{-2}$) above the constant level of emission 
characterising \sgras's quiescent level. Fitting the light curves with a constant, after 
excluding the flaring periods, returns count rates of 0.411, 0.400, 0.339, 
and 0.378~ph~s$^{-1}$, respectively, for the four different \xmm\ observations. 
The extrapolation of the long-term flux evolution of the magnetar, as measured 
by \chandra, as well as the comparison of the new \xmm\ data with archival 
observations, suggests that the magnetar contributes at the level of 
$\sim50$\% to the observed quiescent flux (the observed count rate before 
the magnetar's outburst was 0.196~ph~s$^{-1}$). 

\subsubsection{Can the magnetar be responsible for the flares observed
with \xmm?}

On top of their bright persistent X-ray emission, magnetars show very
peculiar flares on short timescales (from fraction to hundreds of
seconds) emitting a large amount of energy ($10^{37}-10^{46}$\,erg~s$^{-1}$).
They are probably caused by large scale rearrangements of the
surface/magnetospheric field, either accompanied or triggered by
fracturing of the neutron-star crust, as a sort of stellar quakes.
Furthermore, magnetars also show large outbursts where their steady
emission can be enhanced up to $\sim$1000 times its quiescent level
(see Mereghetti 2008; Rea \& Esposito 2010 for recent reviews).  From
the phenomenological point of view, the bursting/flaring events can be
roughly divided in three types: i) {\em short X-ray bursts}, these are
the most common and less energetic magnetar flares. They have short
duration ($\sim0.1-0.2$\,s), and peak luminosity of
$\sim10^{37}-10^{41}$\,erg~s$^{-1}$. They can be observed in a bunch as a
flaring forest, or singularly; ii) {\em Intermediate flares} (in
energy and duration with the flare classes) have typical durations
from a few to hundreds of seconds, and luminosities
$\sim10^{39}-10^{43}$\,erg~s$^{-1}$; iii) {\em Giant flares} are by far the
most energetic Galactic flare ever observed, second only to a possible
Supernovae explosion. The three giant flares detected thus far were
characterised by a very luminous hard peak lasting a bit less than a
second, which decays rapidly into a hundreds of seconds tail modulated
by the magnetar spin period.

Given the vicinity between SGR\,J1745--2900 and Sgr\,A$^{*}$
($\sim2.4$"; Rea et al. 2013), they both lay within the \xmm\
point spread function. Being both flaring sources, we try to use
physical and observational constraints to exclude that the apparent
excess in the flaring activity observed by \xmm\ from the
direction of Sgr\,A$^*$ might be due instead to magnetar flares.

Given the duration and luminosities of the X-ray flares detected by
\xmm\ (see Neilsen et al. 2013), the most similar magnetar flare that we
need to exclude is of the class of {\em intermediate flare}.
Magnetars' intermediate flares are usually observed from young and
highly magnetised members of the class (as it is the case of
SGR\,J1745--2900), either as several consecutive events or singularly
(Israel et al. 2008; Woods et al. 2004). Their spectra are best fit
with a two blackbody model, with temperatures of $kT_1\sim0.5-5$\,keV
and $kT_2\sim6-20$\,keV, from emitting regions of $R_1\sim10-30$\,km
and $R_2\sim0.1-10$\,km, and luminosities of the order of
$10^{38}-10^{42}$\,erg~s$^{-1}$. We then studied in detail the light-curves
and the spectra of our flares. The spectra of all flares were fitted
with a two blackbody model finding a good fit with temperatures of the
order of $kT_1\sim0.7$\,keV and $kT_2\sim6.5$\,keV. Although the
spectral decomposition might resemble that of a typical magnetar
intermediate flare, the derived luminosities of
$\sim10^{35}-10^{36}$\,erg~s$^{-1}$ are low for a magnetar flare.
Furthermore, we find durations around thousands of seconds, which are
also rather long for a magnetar flare.

Even though we cannot distinguish spatially in our data the magnetar
from \sgras, we are confident that the flaring activity we observe
in our \xmm\ observations are not generally consistent with
being due to  SGR\,J1745--2900 and they are produced by \sgras.

\subsubsection{A-posteriori probability of observing the detected flares} 

The observation of four bright or very bright flares in such a short exposure 
($\sim130$~ks) is unprecedented. Following a $\sim3$~Ms \chandra\ monitoring 
campaign, Neilsen et al.\ (2013) estimated \sgras's flaring rate and the 
fluence distribution of the flares. With a total of 39 observed flares, they infer 
a mean flaring rate of $\sim1.2$ flares per 100~ks. 

In particular, we note that only nine bright or very bright flares (according to the 
flare definition in Tab. \ref{DefFlare}) were detected during the 
3~Ms \chandra\ monitoring campaign. Assuming a constant flaring rate,
0.4 such flares were expected during the 133~ks \xmm\ observation, 
as compared to the four that we observed. 

We note that \chandra\ was observing \sgras\ less than 4 hours before 
the start of the \xmm\ observation on August 30, 2014. The red points in 
Fig.\ \ref{XMM} show the \chandra\ light curve, rescaled by a factor 10 for display 
purposes. A weak flare is clearly observed during the $\sim35$~ks exposure. 
An additional $\sim35$~ks \chandra\ observation was performed on October 20,
2014, about one month after the last \xmm\ pointing. A very bright, as well as 
a weak flare were detected during this observation (see Figs.\ 3 and 4, 
Tab. \ref{ExpC2} and Haggard et al.\ 2015). 
Therefore a total of five bright or very bright flares have been observed within 
the 200~ks \xmm\ and \chandra\ monitoring campaign performed at the 
end of 2014, while an average of only 0.6 bright flares would have been 
expected, based on the bright flaring rates previously established (Neilsen et 
al. 2013; 2015; Degenaar et al. 2013). 

Assuming that the flaring events are Poisson distributed and that 
the flaring rate is stationary, we find an {\it a posteriori} probability $P=Poiss_{0-3} 
\left({\lambda}\right)=0.07$~\% of observing four or more bright flares 
during the \xmm\ observations, and a probability 
of $0.04$~\% of observing five or more bright flares in 200~ks. 
These estimates suggest (at just above $3 \sigma$ significance) 
that the observed increase of flaring rate is not the result of stochastic 
fluctuations. Thus, either bright flares tend to cluster, or the flaring 
activity of \sgras\ has indeed increased in late 2014. 

This suggestive change in flaring rate is strengthened by considering 
also \swift\ observations. 
Between 2014 August 30 and the end of the \swift\ visibility window 
(on November 2$^{nd}$), \swift\ observed \sgras\ 70 times for a total 
of 72~ks. On September 10, \swift\ detected a bright flare, 
strengthening the indication of clustering and/or increased flare activity 
during this period (Reynolds et al. 2014; Degenaar et al. 2015)\footnote{While 
\swift\ can detect 
only very bright flares (because of the smaller effective area but similar 
point spread function, compared to \xmm) caught right at their peak (because the 
typical \swift\ exposure is shorter than the flare duration; Degenaar et al. 2013), 
we conservatively consider the same rate of detection as the one observed by \xmm. }. 

Considering all observations of \xmm, \chandra\ and \swift\ 
carried out between mid August and the end of the 2014 \sgras's observability window, 
a total of six bright flares were detected within 272~ks of observations, 
with associated Poissonian probability of $2\times10^{-4}$ (about 
$3.8 \sigma$ significance). 
However, before jumping to any conclusion, we note that the estimated 
probability critically depends on the {\it a-posteriori} choice of the ''start'' 
and duration of the monitoring interval considered. 
To have a more robust estimate of this probability we need to employ 
a well-defined statistic to rigorously identify flares, and then to apply 
a method capable of measuring variations in the flaring rate, without 
making an {\it a-posteriori} choice of the interval under investigation. 
To do so, we perform a Bayesian block analysis. 

\

\section{Bayesian Block analysis}
\label{SBay}

To have a robust characterisation of \sgras's emission we divide all 
the observed light curves into a series of Bayesian blocks (Scargle et al. 2013; 
see also Nowak et al. 2012). 
The algorithm assumes that the light curve can be modeled by a sequence 
of constant rate blocks. A single block characterises light curves in which 
no significant variability is detected. Significant variations will produce blocks 
with significantly different count rates and separated by change points. 
The over-fitting of the light curve is controlled by the use of a downward-sloping 
prior on the number of blocks. 

\subsection{Bayesian block algorithm}

We use the implementation of the ''time tagged'' Bayesian block case described 
by Scargle et al. (2013) and provided by Peter K.G. 
William\footnote{https://github.com/pkgw/pwkit/blob/master/pwkit/\_\_init\_\_.py}. 
The code employs a Monte Carlo derived parameterisation of the prior on the number 
of blocks, which is computed from the probability $p0$, given as an estimation of 
false detection of an extraneous block (typically set at 5~\% Williams et al.\ 2014; 
Scargle et al. 2013). 
The algorithm implements an iterative determination of the best number of blocks 
(using an {\it ad hoc} routine described in Scargle et al. 2013) and bootstrap-based 
determination of uncertainties on the block count rate. This implementation 
starts from the un-binned, filtered \chandra\ event file in FITS format. 
We modified the algorithm to read \xmm\ event files as well. 
The errors and probabilities of false detection presented in this paper 
are derived from independent procedures described in \S~5.1 and 5.5. 

\subsection{Definition of flare, time of the flare, start and stop time, 
duration and fluence}

We define flares as any Bayesian block with count rate significantly different 
from the one(s) describing the quiescent level (we assume that the quiescent 
emission is constant within each observation). The low flaring 
rate typical of \sgras\ allows a good characterisation of the quiescent level 
in all the light curves analysed. Most of the flares are characterised by only 
one flaring block (i.e., a simple rise to a peak value and then a fall back to 
the quiescent level). However, bright or very bright flares can present significant 
substructures generating more than one flaring block for each flare. 
Long bright flares can easily be disentangled from a series of several distinct 
flares, because the latter have a non-flaring block separating the flares\footnote{
We note that, according to this definition, very large amplitude flare substructures 
(where the mean count rate significantly drops to the level observed during quiescence) 
would results in the detection of multiple flares. A similar occurrence has been 
reported by Barriere et al. (2014). In fact, during the \nustar\ observation taken on 
2012 July 21$^{st}$, the authors, through a Bayesian block method, detected two 
flares (J21\_2a and J21\_2b) separated by a short inter-flare period. 
No such events are currently present in the \xmm\ or \chandra\ archives. }. 
For each flare, we define as the flare start and stop time the first and the last 
of the change points characterising the flaring blocks. The flare duration shall be 
the sum of the durations of the flaring blocks. The flare time shall be the 
mid point of the flaring block with the highest count rate. This definition is also 
applied if a flare is in progress either at the start or at the end of the observation. 
We compute the fluence in each flaring block starting from  
the flare count rate during the flaring block, once corrected for pile-up and 
converted to a flux. 
To remove the contribution from background emission, contaminating point 
sources (e.g. SGR~J1745-2900, CXO-GC~J174540.0-290031) and different 
levels of diffuse emission (induced by the different PSF), we subtract 
the quiescent blocks count rate (averaged over all the quiescent blocks of the 
observation under investigation) from the count rate of the flaring block.
We then obtain the fluence of each flaring block by multiplying the ''corrected 
flare'' block count rate by the block duration. 
The total flare fluence shall be the sum of the fluences of all the flaring blocks 
composing the flare (see last column of Tab.\ \ref{ExpC1}, \ref{ExpC2}, \ref{ExpC} 
and \ref{ExpXMM}). 

\subsection{Results}

\begin{figure*}
\includegraphics[height=1.1\textwidth,angle=-90]{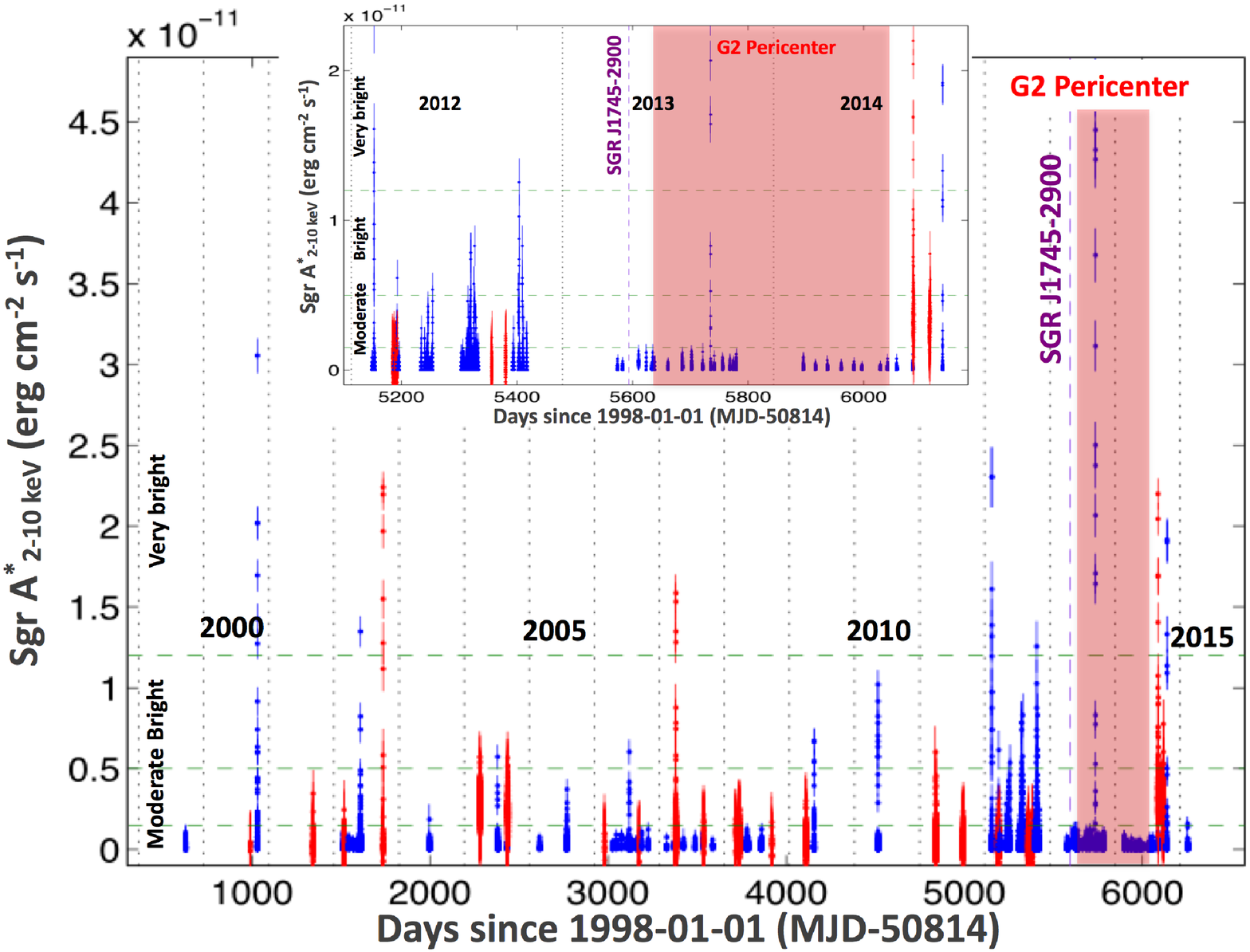}
\caption{{\it (Main figure)} \xmm\ (red) and \chandra\ (blue) coverage of X-ray emission from \sgras. 
Each blue or red point corresponds to a bin of 300~s and shows the flux (with the conversion 
factors shown in Tab.~\ref{Tabpup}) measured by either \chandra\ or \xmm, 
once corrected for the pile-up effect (see \S~\ref{Spileup} and 
Tab. \ref{Tabpup}). 
Flares are manifested as rapid and significant deviations from the quiescent level. 
The quiescent level of emission observed by \xmm\ fluctuates (see red data points) 
because of the contribution from point sources within the extraction region 
chosen for this instrument (see \S~\ref{datared}). 
The dotted lines correspond to the yearly separation. From 2001 to 2005 only 
sporadic observations were performed. In 2006-2008 several observations 
were performed. The long 2012 XVP as well as the 2013-2014 monitoring are 
now allowing good characterisation of the evolution (if any) of the X-ray properties 
over time-scales of a year. The start of the outburst of SGR~J1745-2900 is marked 
with a violet dashed line. The pink box indicates the approximate time of the passage 
of G2 at peri-center. {\it (Upper panel)} Zoom of the last three years of monitoring. 
Several bright-or-very bright flares are observed at the end of 2014, with a higher  
frequency compared to previous observations. 
No moderate or bright flare is observed between the beginning of 2013 and mid 2014, 
in contrast to the frequent occurrence of moderate flares observed in 2012. }
\label{lightcurves}
\end{figure*}
\begin{figure*}
\includegraphics[height=1.1\textwidth,angle=-90]{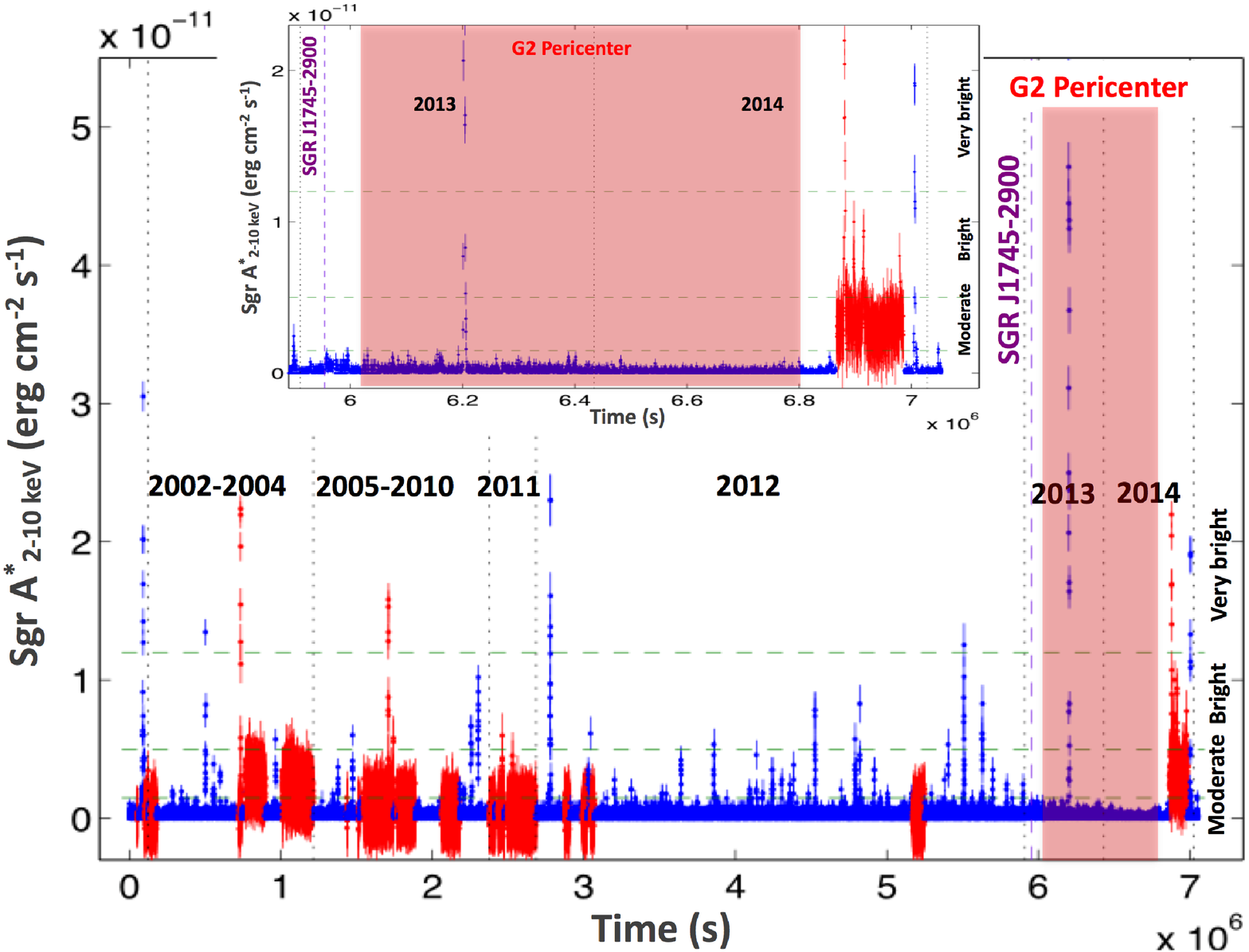}
\caption{{\it (Main figure)} \xmm\ (red) and \chandra\ (blue) light curves of the 2--10~keV flux emitted 
by \sgras. Gaps between observations are removed. Dotted black vertical lines 
separate the different years or periods. The longest total exposure was obtained in 2012. 
The dashed violet vertical line indicates the start of the outburst of the magnetar 
SGR~J1745-2900 on 2013 April 25 (\S~2.1). The pink box indicates the approximate 
time of the peri-center passage of G2 (Gillessen et al. 2013; Witzel et al. 2014). 
The dashed green horizontal lines roughly indicate the demarcation between weak, 
moderate, bright and very bright flares. The different flare types are defined here on 
the basis of their fluence and not by their peak count rates. The thresholds between 
weak and moderate and between moderate and bright flares indicate the average 
count rate for flare lengths of 1~ks.  Longer flares lasting 1.7~ks are considered 
for displaying the threshold between the bright and very bright flares. 
The brightest flare has been detected on 2013 September 14 (Haggard et al. 2015). 
We note that, before 2013, weak, moderate and bright flares are randomly 
distributed within the 15 years of observations. This \sgras\ light curve suggests a lack 
of moderate flares during 2013 and 2014, while we observe a series of 5 bright flares 
clustering at the end of 2014, several months after the peri-center passage of the bulk of 
G2's material. 
{\it (Upper panel)} Zoom on the 2013-2014 period that shows that no moderate flare was 
observed, while 5 bright flares were observed right at the end of the \xmm\ and \chandra\ 
monitoring campaigns. An additional bright flare (not shown here) was detected 
by \swift\ on October 10, 2014, right in the middle of the \xmm\ monitoring campaign. 
These observations suggest an increased flaring rate of \sgras\ during fall 2014.}
\label{Slide1}
\end{figure*}
Figures \ref{lightcurves} and \ref{Slide1} show 
the \chandra\ and \xmm\ light curves of \sgras\ in blue and red, respectively. 
We present here, for the first time, the light curves accumulated since 2013. 
These have been obtained in the course of large \chandra\ and \xmm\ monitoring 
campaigns aimed at studying any variation in the emission properties of \sgras\ induced 
by the close passage of the G2 object (PIs: Haggard; Baganoff; Ponti). 
A detailed study of the possible modulation of the quiescent emission, induced 
by the passage of the cloud, is beyond the scope of this paper and 
will be detailed in another publication. Here we focus our attention on the flaring 
properties only. 
A total of 80 flares have been detected in the period between 1999 and 
2014 (11 by \xmm\ in 1.5~Ms; 20 by \chandra\ between 1999 and 2011, 
in 1.5~Ms; 37 by \chandra\ in 2012, 2.9~Ms; 12 by \chandra\ between 2013 
and 2014, in 0.9~Ms). 
The details of all observations, of all flaring blocks and all flares are reported 
in Tables \ref{ExpC1}, \ref{ExpC2}, \ref{ExpC} and \ref{ExpXMM}. 

The first systematic study of the statistical properties of a large sample of 
\sgras's flares was published by Nielsen et al.\ (2013). The authors 
analysed the 38 \chandra\ HETG observations accumulated in 2012
with a total exposure of $\sim3$~Ms and employed two methods 
to detect flares. Through an automatic Gaussian fitting technique, 
the authors detected 39 flares and provided full details for each 
flare (see Tab. 1 of Neilsen et al.\ 2013). Thirty-three flares are in common. 
We detect five flares, missed by the Gaussian fitting method employed 
by Nielsen et al. (2013). These flares are characterised by low rates 
(in the range $0.01-0.005$~ph~s$^{-1}$) and long durations (lasting 
typically $3-13$~ks), therefore are easily missed by the Gaussian 
method more efficient in detecting narrow-peaked flares. 
On the other hand, our method misses seven flares detected instead by 
Neilsen et al. (2013). Our smaller number of detected flares is a consequence 
of limiting our study to the zeroeth order (ACIS-S), therefore to smaller statistics
(indeed the 7 flares missed, are within the weakest ones detected 
by Neilsen et al.\ (2013), in particular all those having fluences lower 
than 23~photons). 
Neilsen et al. (2013) also employ a different technique to detect flares based 
on a Bayesian block algorithm resulting in the detection of 45 flares. 
The various methods provide consistent results for moderate, bright 
and very bright flares and only differ in the detection of weak flares. 

At first glance, no variation on the flaring rate appears evident 
before and during the peri-center passage of G2. On the other hand, three flares, 
including a very bright one, were detected about 6 months after peri-center 
passage (see Tab. \ref{ExpC}). 

\subsection{Distribution of flare fluences}

\begin{figure}
\includegraphics[height=0.6\textwidth,width=0.45\textwidth,angle=-90]{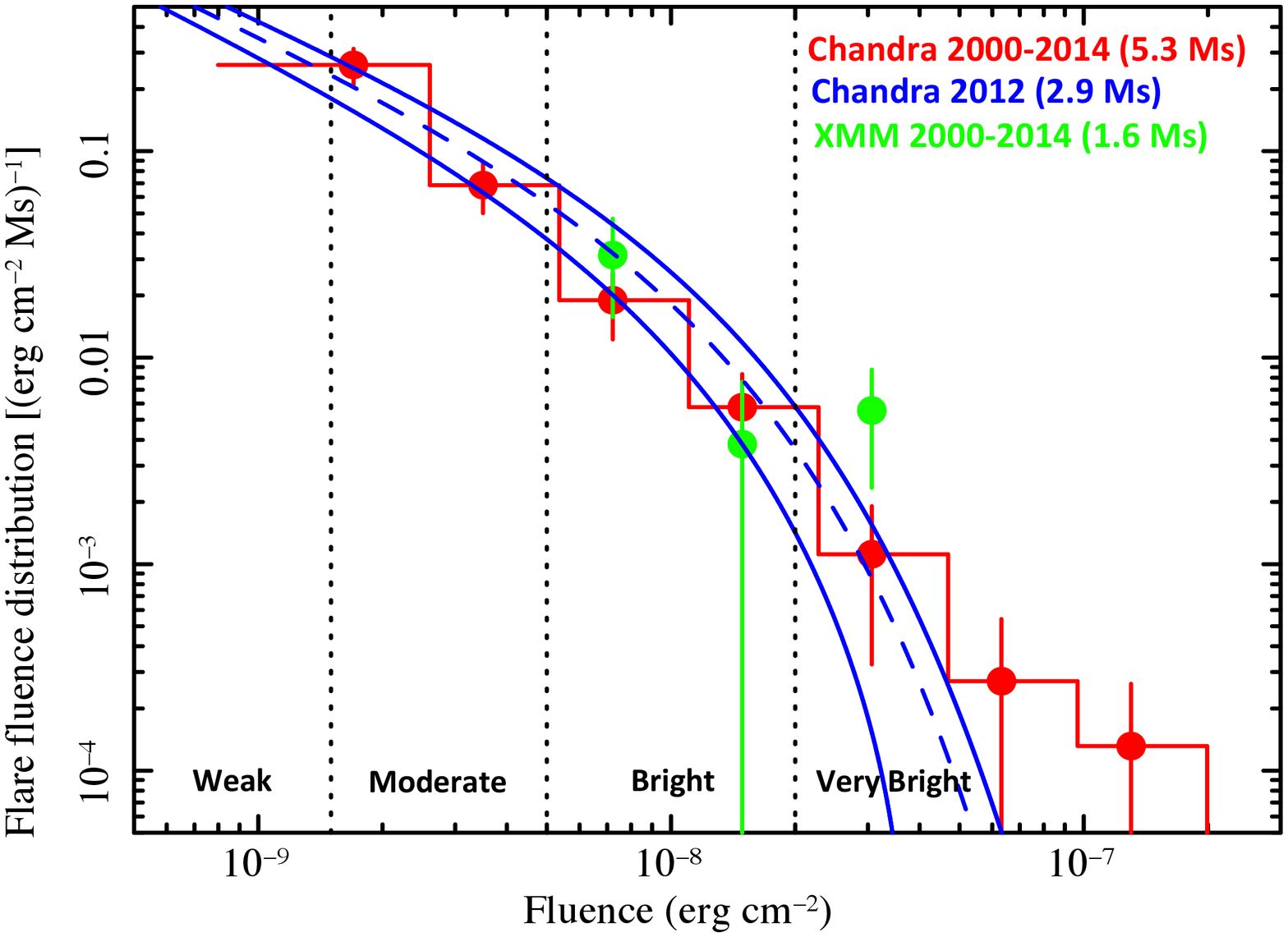}
\caption{Distribution of flare rate as a function of fluence (in erg~cm$^{-2}$~Ms$^{-1}$), 
as observed during 15 years of \chandra\ (red) and \xmm\ (green) monitoring. 
The blue dashed and solid curves indicate the best-fit and 1-$\sigma$ uncertainties 
on the fluence distribution observed during the \chandra\ XVP campaign in 2012 
(Neilsen et al. 2013). The vertical dotted lines indicate the fluence intervals  
characterising the various types of flares (see Tab. \ref{DefFlare}).}
\label{Slide10}
\end{figure}
The red points in Fig. \ref{Slide10} show the distribution of flare fluences 
(normalised to 1~Ms) observed by \chandra\ over the past 15 years, while 
the blue dashed and solid lines show the best fit and 1-$\sigma$ uncertainties 
on the fluence distribution estimated from the \chandra\ XVP campaign in 2012 
(Neilsen et al. 2013). We observe remarkably good agreement between 
the two. In particular, even if we did not correct the flaring rate for 
completeness (particularly important for weak flares), the agreement at low 
fluences indicates that \chandra\ (ACIS-I and ACIS-S with no gratings) 
and \xmm\ are complete in detecting moderate-or-bright and bright-or-very 
bright flares, respectively. 
In addition, we note that both \xmm\ and \chandra\ show a subtle deviation, 
suggesting a higher number of very bright flares is observed during the entire 
dataset analysed, as compared to the 2012 Chandra campaign only. 
This excess might be a result of the inclusion of the latest 2013-2014 campaign. 

\section{Variation of the flaring rate? }

The Bayesian block analysis of the \xmm\ and \chandra\ light curves 
confirms the presence of several bright-or-very bright flares occurring at the 
end of 2014 and allows us to measure the basic flare characteristics. 
To check for any variation of the flaring rate, in an independent 
way from the {\it a posteriori} choice of the start of the interval under investigation 
(see discussion in Section \ref{envelope}), we consider each flare as an event  
and then we apply the Bayesian block method to measure any 
variations of the event rate. 

\subsection{MonteCarlo simulations to estimate the uncertainties 
on flaring rates in a given time bin} 
\label{uncertainties}

We estimate the uncertainty on the number of flares that we expect 
over a given observing time interval, and therefore on the flaring rate, based on Monte Carlo
simulations. The simulations are performed assuming that the flares 
follow the fluence distribution 
as observed during the \chandra\ XVP campaign and reported in Fig. 5 
(blue dashed line, see also Neilsen et al. 2013). 
We first compute the integral of the flare fluence distribution to estimate 
the total number, N$_{tot}$, of flares expected for the entire duration of the 
monitoring (\xmm, \chandra\ and \swift\ 2014). 
Assuming that the flares are randomly and uniformly distributed 
in time we simulated N$_{tot}$ flare occurrence times. 
Then, for each time interval under consideration (i.e. covering the duration of the monitoring 
from a single, or a combination of more observatories), 
we count the number of simulated random occurrences, $N_{sim}$, within that interval. 
We randomize the fluences of the $N_{sim}$ flares within each interval, by drawing $N_{sim}$
random numbers from the \chandra XVP fluence distribution. 
Of these we consider only a given class of flares (e.g. bright and very bright)
 and derive a simulated flare rate associated with this class. 
We repeat this procedure $10^3$~times, and 
estimate the corresponding standard deviation of flare rate. Finally, 
we use this value as the uncertainty associated 
with the observed rate of the given class of flares and within each time interval of interest.

\subsection{\chandra\ observed flaring rate} 
\label{RateCha}

\begin{table*}
\small
\begin{center}
\begin{tabular}{ | l c c c c }
\hline
                                    & \chandra\ 2012 & \chandra\ 1999-2014            & \xmm\ 2000-2014                                        & \xmm\ + \\
                                    &                          &                                               &  [2000-2012] [Aug-Sep 2014]                     & \chandra\ (1999-2014) + \swift\ (2014) \\
\hline
All flares                      & $1.08\pm0.10$ &                                               &                                                                     &  \\
Bright-Very Bright       & $0.26\pm0.05$  & $0.26\pm0.02$                      & $0.45\pm0.16 [0.25\pm0.10] [2.6\pm1.3]$ & $0.27\pm0.01/2.52\pm0.98$\ddag \\
Moderate-Very Bright & $0.97\pm0.12$  & $0.82\pm0.05$                      &                                                                     &  \\
Moderate-Bright         & $0.76\pm0.12$  & $0.69\pm0.07/0.11\pm0.05$\dag &                                                              &  \\ 
Moderate                    & $0.70\pm0.11$  & $0.67\pm0.07/0.11\pm0.05$\dag &                                                              &  \\ 
\hline
\end{tabular}
\caption{Flaring rates (day$^{-1}$) for different types of flares as observed by \chandra\ and \xmm\ 
during different observing intervals. When a variation of the flaring rate is observed, 
the flaring rate of the two blocks are reported. For \xmm\ we report also (in parentheses) 
the flaring rate observed before the end of 2012. 
\dag The change point (variation in the flaring rate) is observed on June 5, 2013. 
\ddag The change point is observed on August 31, 2014.}
\label{TabFlaringRate}
\end{center}
\end{table*} 
We first compute the flaring rate during the 2012 \chandra\ 
observations only ($2.9$~Ms exposure). We observe a rate of all flares (from 
weak to very bright) of $1.08\pm0.10$ per day, and $0.26\pm0.05$ 
bright or very bright flares per day (see Tab. \ref{TabFlaringRate}). 
These values are consistent with the numbers reported by Neilsen et al. (2013; 2015). 

To expand the investigation to \chandra\ observations performed with a different 
observing mode, we henceforth discard the weak flares. Taking all new 
and archival \chandra\ observations from 1999 until the end of 2014 
($\sim5.3$~Ms exposure), we observe that the rate of moderate-to-very-bright 
flares has a mean value of $0.82\pm0.05$ flares per day (a rate of $0.97\pm0.12$ 
was observed during 2012), while the rate of bright-or-very bright flares 
is $0.26\pm0.02$ per day. Restricting this investigation to the 
\chandra\ data only (with $p0<0.05$), no significant difference in the rate of 
total flares is observed. 

Despite the invariance of the total flare rate over the 15 years 
of \chandra\ observations, we note a paucity of moderate flares over the 
2013-2014 period, compared to previous observations (see Fig. \ref{Slide1}). 
Indeed, the rate of moderate flares was $0.67\pm0.07$ per day, showing a tentative 
indication of a drop to $0.11\pm0.05$ after June 5, 2013 (see Fig. \ref{Slide1}; 
$p0=0.08$). 
Moderate flares, if present, would be detected even considering the additional 
contamination induced by SGR~J1745-2900. 
\begin{figure}
\includegraphics[height=0.6\textwidth,width=0.45\textwidth,angle=-90]{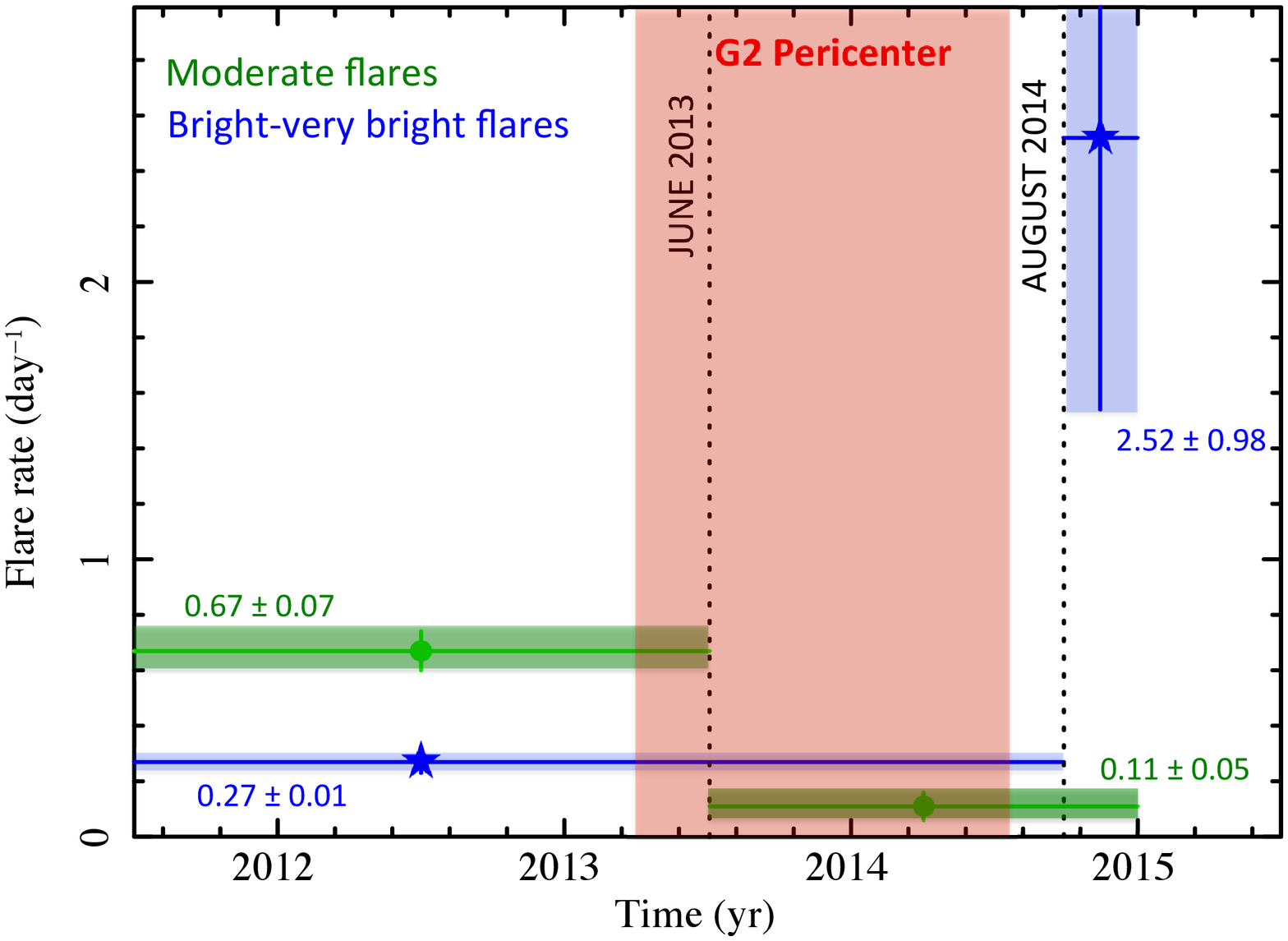}
\caption{Zoom of the variation of the flaring rate of \sgras\ over the past 3.5 years. 
The green data points show the moderate-bright flaring rate (\chandra\ only), 
while the green regions indicate the modulation of the rate and its uncertainty 
($1 \sigma$) derived through Monte Carlo simulations (see \S \ref{uncertainties}). 
The moderate flaring rate is observed to be constant for more than a decade and 
to significantly (at the 96~\% confidence level; see \S~5.5) drop after 2013 June 5, 
several months before the peri-center passage of the bulk of G2's material (see pink box). 
The blue data points show the bright and very bright flaring rate (\chandra\ and \xmm)
as derived from the combined \chandra, \xmm\ and \swift\ observing campaigns 
over the past 15 years, while the shaded light blue regions show 
the uncertainties (see \S~5.1). 
After being constant for more than 14 years, the bright-or-very bright flaring rate 
significantly ($\sim99.9$~\% confidence level) increased after 2014 August 31, 
several months after the peri-center passage of G2. }
\label{Slide9}
\end{figure}

\subsection{Flaring rate observed with \xmm}

The total cleaned exposure of the entire \xmm\ monitoring of \sgras\ 
(from 2000 until the end of 2014) is composed of about $1.6$~Ms of 
observations. We detected eleven flares. This lower number 
compared to \chandra\ can be attributed to the inability of \xmm\ to detect 
weak and moderate flares. 
Eight either bright or very bright flares are detected by \xmm, resulting 
in a mean rate of $0.45\pm0.16$ bright flares per day (see Tab. 3). 
This rate is higher than the one measured by \chandra. This is due to the 4 bright 
or very bright flares detected during the observations accumulated in fall 2014. 
In fact, if only the observations carried out before the end of 2012 are considered, 
the rate of bright or very bright flares drops to $0.25\pm0.10$ per day, 
consistent with the rate derived with \chandra\ and seen with \swift\ 
in 2006-2011 (Degenaar et al. 2013). On the other hand, if we consider only 
the \xmm\ observations carried out in August and September 2014, the 
observed rate is $2.6\pm1.3$ per day. 

\subsection{\xmm, \chandra\ and \swift\ light curves to constrain the change of 
the rate of bright flares}

Combining the light curves of \xmm, \chandra\ and \swift\ (2014) we obtain 
a total cleaned exposure time of $\sim6.9$~Ms. 
During this time, 30 bright-or-very bright flares were detected. 

The Bayesian block analysis now significantly detects a variation in the rate 
of bright or very bright flares in late 2014. In particular, a constant flaring rate, 
from 1999 until summer 2014, of $0.27\pm0.01$ bright flares per day, is found. 
On August 31, 2014 we find a change point in the flaring rate such that the rate 
significantly increased to $2.52\pm0.98$ per day (see Tab. 3), a factor $\sim10$
higher than the prior value (see Fig. \ref{Slide9}). 
This variation of the flaring rate is not detected by the Bayesian block routine 
if we require a value of $p0$ smaller than 0.003\footnote{To check the influence 
of the threshold for background cut on the derived flaring rate, we re-computed 
the bright-or-very bright flaring rate with several thresholds. In particular, if no 
cut is applied, we also detect the two brightest flares (the other weaker features are 
not significantly detected by the Bayesian block routine) observed by B\'elanger 
et al. (2005) during the \xmm\ observations performed in 2004. 
Considering these flares (and the additional exposure time) we derive a 
bright-or-very bright flaring rate of $0.29\pm0.01$~day$^{-1}$, therefore 
consistent with the estimated value. }. 

The point in time when the variation of the flaring rate occurred (change point) 
is quite precise (the typical spacing between the 2014 \xmm\ and \chandra\ 
observations is of the order of $1$~month) and took place several months 
after the bulk of the material of G2 passed peri-center. 
In particular no increase in the flaring rate is observed 6 months 
before (e.g. in 2013) and/or during the peri-center passage (Gillessen et al 2013; 
Witzel et al. 2014). 

\subsection{Significance of the flaring rate change} 

To give a rigorous estimate of the probability of detecting a variation 
in the bright and very bright flaring rate of \sgras, we performed 
MonteCarlo simulations. In the simulations we assumed a constant 
flaring rate, and the fluence distribution observed by \chandra\ in 2012 
(see Fig. 5 and Neilsen et al. 2013). 
The latter was used to derive the expected total number, N, of bright 
and very bright flares in the hypothesis that the flaring rate did 
not change since 2012. Assuming that the flares are randomly and 
uniformly distributed (such that any clustering which would produce 
an increase of flaring rate occurs by chance), we simulated N 
occurrence times for the bright and very bright flares over a total 
exposure which corresponds to the duration of the combined \chandra, 
\xmm\ and \swift\ (2014) campaigns\footnote{Despite some bright or very bright 
flares could be missed by a short ($\sim1$~ks) \swift\ observation, 
the simulations conservatively assume a 100\% efficiency in detecting flares. 
We checked that no bias is introduced by simulating events instead 
than the full X-ray light curves (the threshold for detecting bright or 
very bright flares is much higher than the Poisson flux distribution 
associated with the quiescent emission, therefore no spurious detections 
are induced by the latter) or considering the total fluence distribution, instead 
than the one observed with \chandra\ in 2012 (Neilsen et al. 2013). }. 
We repeated this procedure $10^4$ times, each time applying the 
Bayesian block algorithm (with p0=0.003) to measure how often 
the Bayesian method detects a spurious increase of the flaring rate. 
This happened 10 times out of $10^4$ simulations. 
Therefore the significance of the detected variation in \sgras's 
bright and very bright flaring rate is $\sim99.9$~\% ($\sim3.3 \sigma$). 
The presence of observing gaps does not affect the estimated probability 
(if the flaring rate is constant, e.g., if the flare occurrence times are uniformly 
distributed). 

In the same way, to estimate the significance of the variation of the moderate-or-bright 
flares, we simulated $10^{4}$ light curves with an exposure as observed within 
the \chandra\ campaign. From these we selected the moderate-or-bright flares only, 
then we applied the Bayesian block algorithm with $p0=0.08$, such 
as observed in \S~\ref{RateCha}. We observe that spurious variations happened 
394 times, suggesting a significance of the variation of the moderate flares at 
the $96$~\% confidence level. 

\section{Variation of the total luminosity emitted as bright-or-very bright flares} 

\begin{figure}
\includegraphics[height=0.60\textwidth,width=0.45\textwidth,angle=-90]{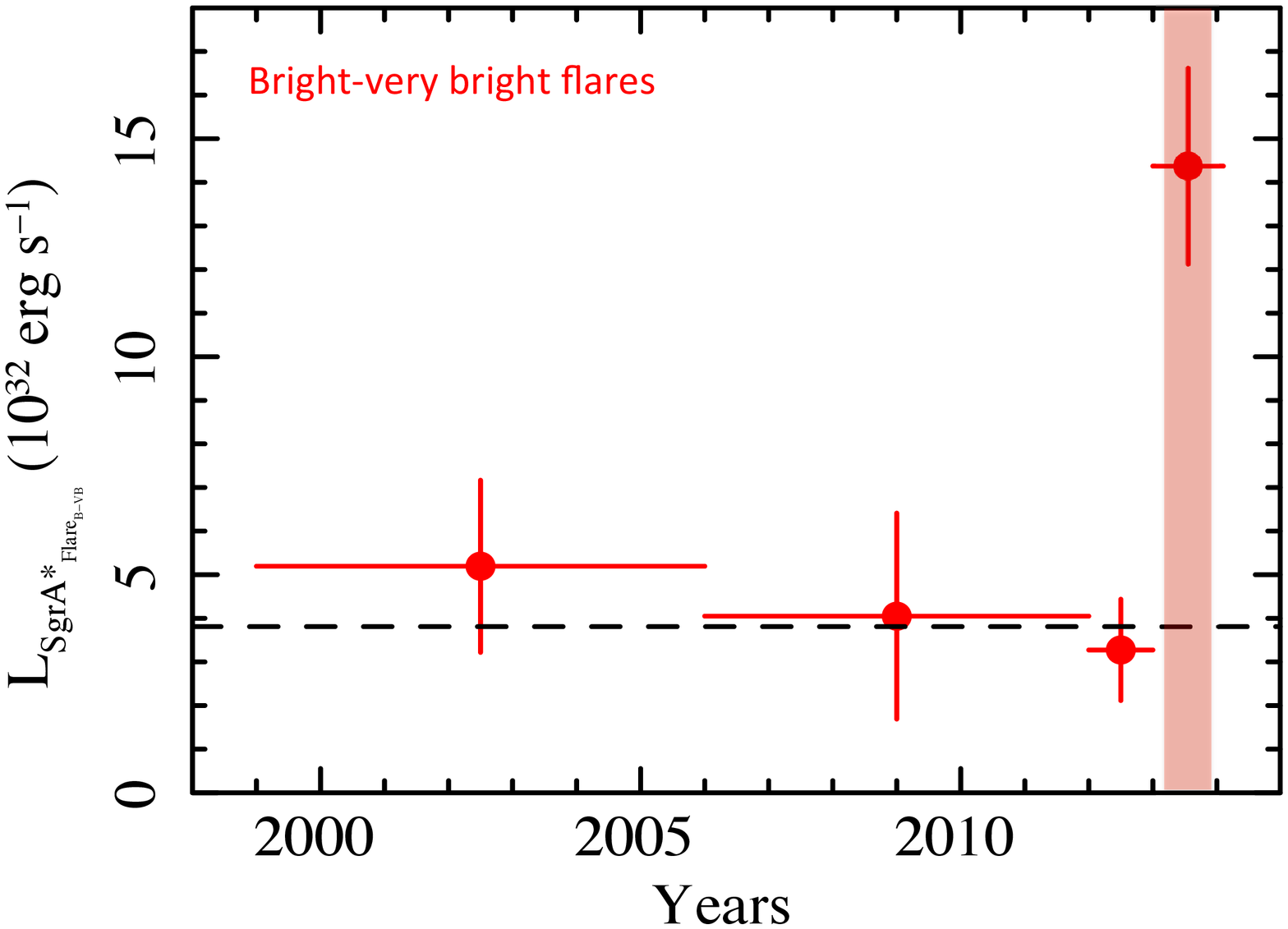}
\caption{Light curve, over the 15 years of \xmm\ and \chandra\ monitoring, 
of the total luminosity emitted by \sgras\ as bright or very bright flares. 
A constant luminosity is observed up to the end of 2012 (see dashed line). 
A significant ($\sim3.5$--$\sigma$ confidence) increase, by a factor $\sim3.7$, is 
observed during the years 2013-2014. Error bars indicate the 1-$\sigma$ 
uncertainty as derived in \S~6. The pink region indicates the period of peri-center 
passage of G2. }
\label{energetics}
\end{figure}
Figure \ref{energetics} shows the light curve, over the past 15 years 
of \xmm\ and \chandra\ monitoring, of the average luminosity emitted 
by \sgras\ in the form of bright-or-very bright flares. 
We choose a single time bin for the long \xmm\ and \chandra\ 
exposure in 2012 and one for the 2013-2014 campaign, while we 
divide the historical monitoring from 1999 to 2011 in two time bins, 
having roughly similar exposures. 
The amplitude of the uncertainty on the measurement of the energy released 
by \sgras, in the form of flares, depends both on the uncertainty on the 
measurement of the energetics associated with each single flare and on the 
uncertainty on the number of flares that we expect in the given interval. 
The first one can be estimated through error propagation and it is typically 
negligible compared to the second, given the flare distribution and the 
intervals considered here. We estimate the uncertainty on the total 
luminosity in flares over a given interval through the same procedure as 
described in section \ref{uncertainties}.

\sgras\ shows an average luminosity in bright-or-very bright flares of 
$L_{\rm Fx_{BVB}}\sim3.8\times10^{32}$~erg~s$^{-1}$ (assuming a 8~kpc 
distance) over the 1999-2012 period (see Fig. \ref{energetics}). 
No significant variation is observed. On the other hand, a significant increase, 
compared to a constant ($\sim3.5 \sigma$ confidence; $\Delta\chi^2=19.6$ for 3 dof), 
of a factor of 2-3, in \sgras's luminosity is observed over the 2013-2014 period (see 
Fig. \ref{energetics}). This result, such as the variation of the flaring rate, 
suggest a change in \sgras's flaring properties. 

\subsection{Historical and new fluence distribution}

\begin{figure}
\includegraphics[height=0.6\textwidth,width=0.45\textwidth,angle=-90]{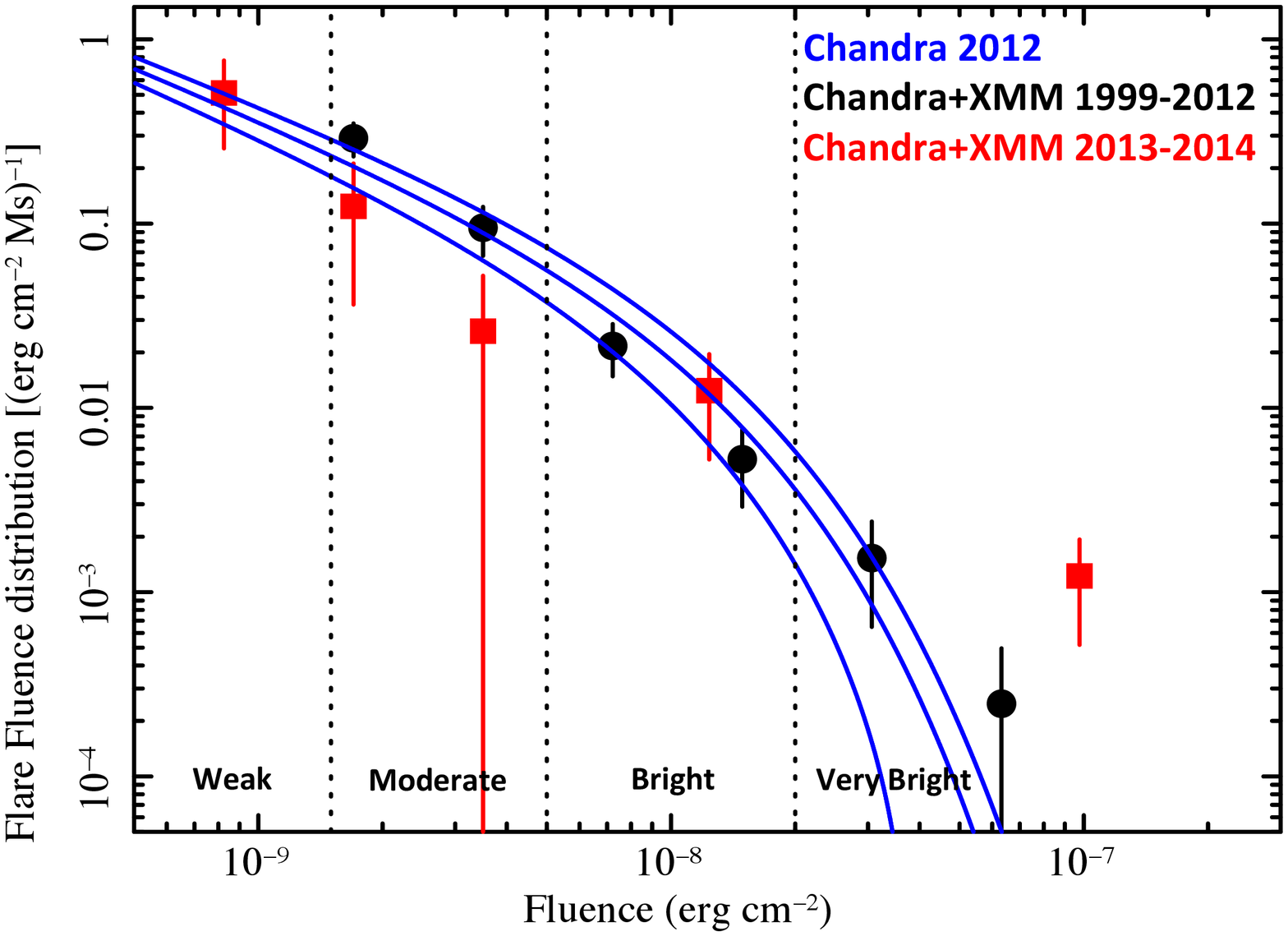}
\caption{Change of the distribution of the flare rate as a function of fluence (in 
erg~cm$^{-2}$~Ms$^{-1}$), as observed during 15 years of \chandra\ and \xmm\ monitoring. 
The flare fluence distribution observed during the 1999-2012 period (black circles) is 
consistent with that observed during the XVP campaign (blue lines). 
A variation of the flare fluence distribution has been observed during the past two years 
(red squares). In particular, both a larger number of very bright flares (and possibly fewer 
moderate flares) have been observed. The dotted vertical lines indicate the different types 
of flares. 1--$\sigma$ uncertainties are displayed. }
\label{EvolutionFluence}
\end{figure}
The black circles in Fig. \ref{EvolutionFluence} show the flare fluence distribution as 
observed during the 1999-2012 period. The \xmm\ and \chandra\ monitoring campaigns, 
lasting for more than a decade show no significant variation of \sgras's flaring activity, 
compared to that observed in 2012 (Neilsen et al. 2013). 
On the other hand, during the past two years a significant change in the flare fluence 
distribution is observed. In fact, during the 2013-2014 period (see red squares in Fig. 
\ref{EvolutionFluence}) the fluence distribution deviates from the one observed in 2012. 
In particular, we observe a clear increase in the number of very bright flares and 
the tentative detection of a decrease in the number of moderate flares, during the 
past two years. These results confirm the conclusions obtained through the study of 
the variation of the flaring rate and luminosity. 
Figure \ref{EvolutionFluence} shows that, despite the tentative detection of 
a decrease in the rate of moderate flares, the enhanced 
rate of very bright flares drives the observed increase of the total energy released 
by \sgras. 

\section{Discussion}

Through a Bayesian block analysis of \sgras's flaring rate light curve, we observe a 
$\sim99.9$~\% significance increase of the rate of bright or very bright flare 
production, from $0.27\pm0.04$ to $2.5\pm1.0$ day$^{-1}$, 
starting after summer 2014. We also observe a tentative 
detection ($\sim96$~\% significance) of a decrease in the rate of 
moderate-bright flares since mid 2013 (see Fig. 8 and \ref{Slide9}). 
Despite the decrease in the rate of moderate flares, the total energy emitted 
by \sgras, in the form of bright-or-very bright flares, increased (at $\sim3.5 \sigma$
confidence level) over the 2013 and 2014 period, compared to historical 
observations. The close time coincidence with the passage of G2 at 
peri-center is suggestive of a possible physical connection. 
Yet, since the power spectrum of the regular X-ray flaring activity is not well 
known and because of the frequent monitoring in this specific time period, 
one cannot exclude that the observed variation is a random event.

Is the observed flaring activity of \sgras\ an odd behaviour of a peculiar source? 
Has a similar change of the flaring rate ever been observed in other sources? 

\subsection{Comparison with quiescent stellar mass BH systems}

Flaring activity is a common characteristic of the optical and infrared counterparts of compact 
binaries (see e.g. Bruch 1992), including quiescent stellar-mass BHs in X-ray binaries. 
In particular, Zurita, Casares \& Shahbaz (2003) presented a detailed study of this topic, 
showing that intense flaring -- amplitudes varing from 0.06 to 0.6 optical magnitudes -- is 
systematically present in the light curves of four BH systems and one neutron star X-ray binary. 
These flares are bright enough to rule out a companion star origin, and therefore 
are believed to arise from the accretion flow. 
Flare durations range from a few hours down to the shortest sampling performed, typically 
of the order of minutes. The optical power density spectra show more power at low 
frequencies and can be modelled by a power-law with a spectral index of $\beta\sim$ --1 
for three of the systems, and $\beta \sim$ --0.3 for the other, hence closer to white noise. 
High cadence observations of XTE J118+40; (Shahbaz et al. 2005) have shown that 
this power-law extends up to frequencies corresponding to time scales of at least a few 
seconds. The number of flares is witnessed to decrease with duration and long events 
tend to be brighter. 

Interestingly, the long-term activity is known to vary in at least one of the BHs 
analysed (A0620--00), where active and passive quiescent states have been 
reported by Cantrell et al. (2008). The former are slightly brighter and bluer, 
whereas much weaker flaring activity is observed during the latter, when the 
light-curve is largely dominated by the modulation produced by the Roche-lobe 
shaped companion star. Typical time scales for these states go from a month to years, 
but transitions seem to occur in a few days.  

Time resolved X-ray observations during quiescence have been also performed 
for the X-ray brightest of the above objects, V404~Cyg, which has a quiescent 
X-ray luminosity of $\sim10^{34}$ erg s$^{-1}$ (see e.g. Hynes et al. 2004). 
We note that this is a factor $10^{2-4}$ brighter than typically observed 
for stellar-mass black holes in quiescence (see e.g. fig. 3 in Armas Padilla et 
al. 2014). Numerous X-ray flares with an amplitude of one order of magnitude 
in flux  are observed, and they are correlated with optical events, with short-time 
time-lags that can be explained by X-ray (or Extreme Ultraviolet) reprocessing. 
X-ray flares have also been detected in some quiescent X-ray binaries with 
neutron stars accretors (e.g. Degenaar \& Wijnands 2012).

This evidence is in line with the idea that flaring activity is not a peculiarity 
of \sgras\ (e.g. related to a process unique to supermassive BH environments), 
instead it appears to be a common property of quiescent BH (e.g. possibly related 
to the accretion flow). In particular, in V404~Cyg, the system with the brightest 
optical and X-ray quiescent level (therefore allowing the best multi-wavelength 
characterisation of the flaring activity) correlated optical and X-ray flares are 
observed (Hynes et al. 2004). In addition, no lag between the X-ray and optical 
variations are observed, implying that they are casually connected 
on short time-scales. This indicates that the optical and X-ray flaring activity 
are linked by a process generated in the inner accretion flow. 
Furthermore, the observed change in \sgras's flaring properties also appears 
compatible with the long term evolution of the flaring activity in binaries. 
However, we note that if flare durations linearly scale with BH mass, 
it is not possible to detail the flares of BH binaries on the same scaled 
times observed in \sgras. Further analysis, which is beyond the scope 
of this work, is needed to actually probe that the same physical process 
is at work in both type of objects. 

\subsection{Clustering of bright flares: is it an intrinsic property of the underlying noise process?} 

The origin of \sgras's flaring activity is still not completely 
understood. The observation of bright flares is generally associated 
with the detection of few other weaker flaring events, suggesting that 
in general flares tend to happen in clusters (Porquet et al. 2008; 
Nowak et al. 2012). If so, the observed variation might be the manifestation 
of an underlying process that, although being stationary (showing the same 
average flaring properties on decades time-scales), produces flares not 
uniformly distributed in time. 

Though uninterrupted \xmm\ observations have already suggested a possible 
clustering of bright flares, only now we can significantly show that the 
distribution in time of bright flares is not uniform. 
\sgras's flaring activity, in the IR and sub-mm bands, is dominated by a red noise 
process at high frequencies, breaking at timescales longer than a fraction of a day 
(Do et al. 2009; Meyer et al. 2009; Dodds-Eden et al. 2011; 
Witzel et al. 2012; Dexter et al. 2014 Hora et al. 2014). 
\sgras's X-ray emission appears to be dominated by two different 
processes, one diffuse and constant, dominating during the quiescent periods, 
and one producing narrow, high-amplitude spikes, the so called flares. 
The power spectral shape of the latter is not clearly determined, however the
X-ray light curves appear fairly different from the red-noise dominated light 
curves of AGN with comparable BH mass (Uttley et al. 2002; McHardy et al. 2006; 
Ponti et al. 2010; 2012). 

A correlation is observed between the bright IR flux excursions and the 
X-ray flaring activity (although IR flares typically have longer durations), 
suggesting a deep link between the variability process in these two bands.
X-ray flares might occur only during the brightest IR excursions, therefore 
happening more frequently when the mean IR flux results to be high because 
modulated by the intrinsic red-pink noise process in IR. 
If so, the observed variation in the X-ray flaring rate might suggest the presence 
of long term trends (on timescales longer than the typical X-ray observations) 
generally associated with red or pink (flicker) noise processes. 
In this case, the observed clustering would be an intrinsic property of the noise, 
and the variation would be detected significantly simply because of the increased 
frequency in monitoring of \sgras, without {\it a priori} physical connection with 
the peri-center passage of G2. If this is indeed the case, we would expect 
that the next set of observations will show the same flaring rate as recorded 
in historical data, with another clustering event happening at some 
other random point in the future. 

This is also in line with what is observed in other quiescent BH. Indeed, strong 
flaring activity has been observed in all quiescent stellar mass BH where this 
measurement was possible (Zurita et al. 2003). In particular, at least one object 
(the best monitored, A0620-00) showed a dramatic change in the flaring properties, with 
periods devoid of any observed flare followed by intense flaring episodes. 
Transitions between flaring and non-flaring periods occurred at irregular intervals 
of time-scales typically longer than a (few) day(s). 
Assuming that these time-scales vary linearly with the BH mass, we would expect 
to observe macroscopic changes in the flaring rate to occur on time-scales 
longer than many hundred years, or even longer period (and luminosity 
amplitudes smaller) than what is probed by GC molecular clouds 
(e.g. Sunyaev et al. 1993; Ponti et al. 2010; 2013; Zhang et al. 2015). 

\subsection{Enhanced flaring activity induced by G2?}

An alternative possibility is that the increased flaring activity is induced by the 
passage of G2. Indeed, part of the G2 object could have been deposited close to peri-center, 
generating an increased supply of accreting material in the close environment 
of the supermassive BH and possibly perturbing the magnetic field structure. 
It was predicted and it has been observed that the material composing 
G2 have been stretched close to peri-center, with the bulk of the 
material reaching peri-center, at $\sim10^3$~r$_g$ from \sgras, in early 2014 
and with the full event lasting about one year (e.g., gas at post peri-center 
was already observed in April 2013; Gillessen et al. 2013; Witzel et al. 2014).

When discovered, G2's impact parameter was smaller than its size 
(however, at peri-center the diffuse part of G2 has been highly stretched, 
therefore reducing the impact parameter). Irrespective of the nature of 
G2 (i.e. whether or not it contains a star), 
it is highly likely that part of G2's outer envelope has been detached from 
G2's main body and captured by \sgras's gravitational potential. 
If, indeed, part of G2's material has been left behind and it is now starting 
to interact with the hot accretion flow in \sgras's close environment 
(e.g. through shocks), it could destabilise it, changing the physical conditions 
there, and possibly inducing enhanced accretion. 
Much theoretical work has been performed to predict the evolution of \sgras's 
emission and envision the importance of magnetic phenomena (Burkert et al. 2012; 
Schartmann et al. 2012; Anninos et al.  2012; S{\c a}dowski et al. 2013; 
Ballone et al. 2013; Abarca et al. 2014; De Colle et al. 2014). 
The most likely scenario is that of an increase in the mass accretion rate onto 
\sgras. 
Simulations are, however, limited in the power of resolving the physics 
of the accretion flow down to \sgras's last stable orbit, and therefore to predict 
under what form and extent this will be transformed into radiation. 
It has been predicted that this might result into a slow increase of the quiescent 
emission, of a factor of a few compared to quiescence, on a timescale 
of years or decades. However it is not excluded that the increased accretion rate 
results into enhanced luminosity and rate of the flares. 
For example, it is possible that bright flares might be generated either if the 
material of G2 is very clumpy and remains cold, creating instabilities in the hot 
flow around \sgras\ that would generate accretion in bunches, or by the arrival 
of shocks, induced by G2, onto the BH close environment. 
In particular, based on their accretion-based model, Mo\'scibrodzka et al. (2012) 
estimate that the X-ray luminosity scales as the third power of the mass accretion 
rate ($L_X\sim \dot{M}^{3}$). If so, a small increase of the mass accretion rate 
might rescale, as observed, the X-ray light curve causing an increase of the total 
flare luminosity and frequency. 
On the other hand, since the average X-ray luminosity from close to the BH 
is significantly smaller than the observed quiescent level (Neilsen et al. 2012), 
such change would be less evident in quiescence. 

Although this is indeed possible, we note that another source, G1 (Cl{\'e}net et 
al. 2004; 2005; Ghez et al. 2005), having a large envelope of dust and gas 
with a size larger than its impact parameter and with physical parameters 
very similar to the ones of G2, arrived at peri-center in 2001 (Pfuhl et al. 2015). 
However, no increased flaring rate was registered at that time (see Fig. 
\ref{lightcurves})\footnote{Although we acknowledge that the Galactic center 
has been much more sparsely monitored after the G1 encounter than 
it has been after the G2 passage. }. 

We remind the reader that at $\sim10^3$~r$_g$ the dynamical and viscous time-scale 
correspond to $t_d\sim0.1$~yr and $t_v\sim10$~yr, respectively. 
A viscous timescale of the order of few years suggests that it might be too early 
to observe an increase in the average accretion rate. 
What is clear is that, if the observed variation in the flaring rate has anything to do 
with the passage of G2, this process should not stop within the next few 
months, but it should continue at least on a dynamical-viscous time-scale. 
The coming X-ray observation of \sgras\ will clarify this. 

\section{Conclusions}

\begin{itemize}

\item{} We present the light curves of all \xmm\ and \chandra\ observations of \sgras, 
to characterise its flaring activity. Through a Bayesian block routine we detect and 
describe a total of 80 flares. 

\item{} We also present the new \xmm\ (263~ks) and \chandra\ (915~ks) data 
from the \sgras-G2 X-ray monitoring campaigns. A total of 16 flares, of which 
seven bright-or-very bright ones, have been detected in $\sim1.14$~Ms exposure. 
Apart from a tentative detection ($\sim96$~\% significance) of a slight drop in 
the rate of moderate flares, since June 2013, no other variations 
(frequency or intensity) of the flaring activity is observed before and during 
the peri-center passage of the G2 dust enshrouded object. 

\item{} The last set of \xmm, \chandra\ and \swift\ (2014) observations, 
obtained between August 30 and October 20 revealed a series of six bright flares 
within 272~ks, while an average of only 0.8 bright flares was expected. 
A Bayesian block analysis of the light curve of the bright flares shows a 
significant ($\sim99.9$~\% confidence) variation of the bright flaring rate 
occurring after August 31, 2014. It is of note that we observe that this 
transition happened several months after the peri-center passage of 
the bulk of G2's material (at $\sim10^3$~r$_g$ from \sgras). 

\item{} We also observe a significant ($\sim3.5 \sigma$ confidence) 
increase, by a factor of $\sim3.7$, of the mean luminosity of \sgras\ in bright-or-very 
bright flares, occurring within the past two years (2013-2014). 

\item{} The flaring rate changes are also detected by comparison of the flare fluence 
distribution observed by \xmm\ and \chandra\ to the one observed by \chandra\ in 2012. 

\item{} We note that \sgras's flaring activity is not an atypical behaviour of 
a peculiar BH. Instead, quiescent stellar mass BH show significant 
optical flaring activity (Zurita et al. 2003) and, in at least one case, major 
evolutions of the quiescent activity has been observed (Cantrell et al. 2008). 

\item{} It is not clear whether the observed clustering (increased rate) of 
bright and very bright flares is due a stationary property of the underlying 
variability process or whether we are witnessing the first signs of the passage 
of G2 onto \sgras's close environment. Further X-ray observations might help 
to disentangle between these two hypotheses. 

\end{itemize}

\begin{table*}
\scriptsize
\begin{center}
\begin{tabular}{ | c c r r c c l l l r c r r }
\hline
OBSID &Obs& Exp & Clean        & Obs            & Nb - Nf & Nb & Block & Block & Dur & Observed & Block          & Flare \\
            &mode&       &  Exp           & date            &             &       &Start  & Stop  &        & Count rate & Fluence     & Fluence \\
            &         &       &                  &                    &             &       &          &          &       &                    & 10$^{-10}$          & 10$^{-10}$ \\
            &         &(ks) & (ks)           &                    &             &       &(MJD)&(MJD) & (s) &(ph~s$^{-1}$)& (erg~cm$^{-2}$) & (erg~cm$^{-2}$) \\
\hline
\multicolumn{5}{l}{{\bf 1999}} \\
242   &I&  50.0 &  45.92 & 1999-09-21 02:41:56 & 1 - 0 \\
\multicolumn{5}{l}{{\bf 2000}} \\
1561  &I&  50.0 &  49.3  & 2000-10-26 19:07:00 & 5 - 1 & 2 &  51844.1035 & 51844.2055 & 8810.6 & $0.020\pm0.005$ & 55.5 & \\
      &    &   &        &                     &       & 3 & 51844.2055 & 51844.2444 & 3360.9 & $0.14\pm0.02$     & 373.3 & \\
      &    &   &        &                     &       & 4 & 51844.2444 & 51844.2799 & 3070.7 & $0.069\pm0.007$ & 99.8 & {\bf 528.6} \\
\multicolumn{5}{l}{{\bf 2002}} \\
2951  &I&  12.5 &  12.37 & 2002-02-19 14:26:28 & 1 - 0 \\
2952  &I&  12.5 &  12.37 & 2002-03-23 12:24:00 & 1 - 0 \\
2953  &I&  12.5 &  11.59 & 2002-04-19 10:58:39 & 1 - 0 \\
2954  &I&  12.5 &  12.45 & 2002-05-07 09:24:03 & 1 - 0 \\
2943  &I&  38.5 &  37.68 & 2002-05-22 23:18:38 & 1 - 0 \\	
3663  &I&  40.0 &  37.96 & 2002-05-24 11:49:10 & 3 - 1 & 2 & 52418.7957 & 52418.8506 & 4734 & $0.021\pm0.003$ & 25.6 & {\bf 25.6} \\	
3392  &I& 170.0 & 166.69 & 2002-05-25 15:14:59 & 7 - 3 & 2 & 52420.1699 & 52420.205  & 3032 & $0.022\pm0.004$ & 17.1 & {\bf 17.1} \\
      &    &   &        &                     &       & 4 & 52420.568  & 52420.6255 & 4971 & $0.012\pm0.004$ & 12.3 & {\bf 12.3} \\
      &    &   &        &                     &       & 6 & 52421.2313 & 52421.2435 & 1054 & $0.028\pm0.013$ & 8.2 & {\bf 8.2} \\
3393  &I& 170.0 & 158.03 & 2002-05-28 05:33:40 & 8 - 3 & 2 & 52422.6317 & 52422.6471 & 1331 & $0.050\pm0.016$ & 21.6  \\
      &    &   &        &                     &       & 3 & 52422.6471 & 52422.6698 & 1959 & $0.10\pm0.02$ & 83.2 & {\bf 104.8} \\
      &    &   &        &                     &       & 5 & 52423.2364 & 52423.2954 & 5095 & $0.031\pm0.005$ & 47.2 & {\bf 47.2} \\
      &    &   &        &                     &       & 7 & 52423.7761 & 52423.7881 & 1033 & $0.057\pm0.009$ & 20.1 & {\bf 20.1} \\
3665  &I& 100.0 &  89.92 & 2002-06-03 01:23:33 & 1 - 0 \\	
\multicolumn{5}{l}{{\bf 2003}} \\
3549  &I&  25.0 &  24.79 & 2003-06-19 18:27:51 & 1 - 0 \\
\multicolumn{5}{l}{{\bf 2004}} \\
4683  &I&  50.0 &  49.52 & 2004-07-05 22:32:07 & 1 - 0 \\
4684  &I&  50.0 &  49.53 & 2004-07-06 22:28:54 & 4 - 1 & 2 & 53193.1378 & 53193.1556 & 1530 & $0.071\pm0.016$ & 37.9 \\
      &   &    &        &                     &       & 3 & 53193.1556 & 53193.1691 & 1174 & $0.017\pm0.009$ & 4.6 & {\bf 42.5}\\
5360  &I&   5.0 &   5.11 & 2004-08-28 12:02:55 & 1 - 0 \\ 	
\multicolumn{5}{l}{{\bf 2005}} \\                         
6113  &I&   5.0 &   4.86 & 2005-02-27 06:25:01 & 1 - 0 \\ 	
5950  &I&  49.0 &  48.53 & 2005-07-24 19:57:24 & 1 - 0 \\ 	
5951  &I&  49.0 &  44.59 & 2005-07-27 19:07:13 & 1 - 0  \\ 	
5952  &I&  49.0 &  45.33 & 2005-07-29 19:50:07 & 3 - 1 & 2 & 53581.1053 & 53581.1457 & 3492 & $0.021\pm0.003$ & 18.5 & {\bf 18.5} \\
5953  &I&  49.0 &  45.36 & 2005-07-30 19:37:27 & 3 - 1 & 2 & 53581.9263 & 53581.9511 & 2146 & $0.046\pm0.008$ & 31.5 & {\bf 31.5} \\
5954  &I&  19.0 &  17.85 & 2005-08-01 20:15:01 & 1 - 0 \\
\multicolumn{5}{l}{{\bf 2006}} \\
6639  &I&   5.0 &   4.49 & 2006-04-11 05:32:15 & 1 - 0 \\
6640  &I&   5.0 &   5.1  & 2006-05-03 22:25:22 & 1 - 0 \\ 
6641  &I&   5.0 &   5.06 & 2006-06-01 16:06:47 & 1 - 0 \\	
6642  &I&   5.0 &   5.11 & 2006-07-04 11:00:30 & 1 - 0 \\ 	
6363  &I&  30.0 &  29.76 & 2006-07-17 03:57:24 & 4 - 1 & 2 & 53933.2439 & 53933.2677 & 2059 & $0.07\pm0.03$ & 50.1 \\
      &    &   &        &                     &       & 3 & 53193.1556 & 53193.1691 & 1174 & $0.017\pm0.009$ & 4.6 & {\bf 54.7} \\
6643  &I&   5.0 &   4.98 & 2006-07-30 14:29:21 & 1 - 0 \\
6644  &I&   5.0 &   4.98 & 2006-08-22 05:53:30 & 1 - 0 \\
6645  &I&   5.0 &   5.11 & 2006-09-25 13:49:30 & 1 - 0 \\
6646  &I&   5.0 &   5.1  & 2006-10-29 03:27:16  & 2 - 1 & 1 & 54037.1559 & 54037.1589 & 261 & $0.031\pm0.017$ & 2.2 & {\bf 2.2} \\ 
\multicolumn{5}{l}{{\bf 2007}} \\
7554  &I&   5.0 &   5.08 & 2007-02-11 06:15:50 & 1 - 0 \\ 	
7555  &I&   5.0 &   5.08 & 2007-03-25 22:55:03 & 1 - 0 \\
7556  &I&   5.0 &   4.98 & 2007-05-17 01:03:59 & 1 - 0 \\	
7557  &I&   5.0 &   4.98 & 2007-07-20 02:25:57 & 1 - 0 \\
7558  &I&   5.0 &   4.98 & 2007-09-02 20:18:36 & 1 - 0 \\
7559  &I&   5.0 &   5.01 & 2007-10-26 10:02:59 & 1 - 0 \\
\multicolumn{5}{l}{{\bf 2008}} \\
9169  &I&  29.0 &  27.6  & 2008-05-05 03:52:11 & 3 - 1 & 2 & 54591.4414 & 54591.4424 &  89 & $0.07\pm0.04$ & 2.1 & {\bf 2.1} \\
9170  &I&  29.0 &  26.8  & 2008-05-06 02:59:25 & 1 - 0 \\
9171  &I&  29.0 &  27.69 & 2008-05-10 03:16:58 & 1 - 0 \\	
9172  &I&  29.0 &  27.44 & 2008-05-11 03:35:41 & 1 - 0 \\
9174  &I&  29.0 &  28.81 & 2008-07-25 21:49:45 & 1 - 0 \\		
9173  &I&  29.0 &  27.77 & 2008-07-26 21:19:45 & 1 - 0 \\	
\multicolumn{5}{l}{{\bf 2009}} \\
10556 &I& 119.7 & 112.55 & 2009-05-18 02:18:53 & 8 - 3.5 & 1 & 54969.1108 & 54969.1246 & 1198 & $0.0409\pm0.009$ & 15.2 & {\bf 15.2} \\ 
      &    &   &        &                     &      & 3 & 54969.4034 & 54969.445  & 3598 & $0.024\pm0.005$ & 22.8 & {\bf 22.8} \\ 
      &    &   &        &                     &      & 5 & 54969.9604 & 54969.9805 & 1732 & $0.090\pm0.008$ & 60.7 & {\bf 60.7} \\ 
      &    &   &        &                     &      & 7 & 54970.0357 & 54970.0426 & 589 & $0.0815\pm0.017$ & 17.9 & {\bf 17.9} \\
\multicolumn{5}{l}{{\bf 2010}} \\
11843 &I&  79.8 &  78.93 & 2010-05-13 02:11:28 & 3 - 1 & 2 & 55329.1491 & 55329.1902 & 3556 & $0.104\pm0.006$ & 156.4 & {\bf 156.4} \\
\multicolumn{5}{l}{{\bf 2011}} \\
13016 &I&  18.0 &  17.83 & 2011-03-29 10:29:03 & 2 - 0.5 & 1 & 55649.4479 & 55649.4826 & 2998 & $0.014\pm0.005$ & 10.1 & {\bf 10.1} \\
13017 &I&  18.0 &  17.83 & 2011-03-31 10:29:03 & 1 - 0 \\
\hline
\end{tabular}
\caption{List of all \chandra\ observations on \sgras\ performed between 1999 and 2011. 
For each observation the observation ID, the instrument setup (I, SH and S indicate that the 
observation was taken with the ACIS-I, the ACIS-S with HETG and ACIS-S with no gratings 
and 1/8 subarray, respectively), the total and the cleaned exposure time, the observation start 
date and the number of different blocks detected in our Bayesian block analysis plus the number 
of flaring blocks, are reported. In the following columns we report information about each flaring block. 
We report the block number, the block start and stop times (in MJD), the duration (in seconds), 
the mean observed count rate (in units of ph~s$^{-1}$, with associated error (1-$\sigma$), 
computed through bootstrap simulations) and the fluence (in units of $10^{-10}$~erg~cm$^{-2}$, 
after subtraction of the average contribution of the non flaring blocks). 
The last column shows, for each flare, the total flare fluence in erg~cm$^{-2}$. 
}
\label{ExpC1}
\end{center}
\end{table*} 

\begin{table*}
\scriptsize
\begin{center}
\begin{tabular}{ | c c r r c c l l l r c r r }
\hline
OBSID &Obs& Exp & Clean        & Obs            & Nb - Nf & Nb & Block & Block & Dur & Observed & Block          & Flare \\
            &mode&       &  Exp           & date            &             &       &Start  & Stop  &        & Count rate & Fluence     & Fluence \\
            &         &       &                  &                    &             &       &          &          &       &                    & 10$^{-10}$          & 10$^{-10}$ \\
            &         &(ks) & (ks)           &                    &             &       &(MJD)&(MJD) & (s) &(ph~s$^{-1}$)& (erg~cm$^{-2}$) & (erg~cm$^{-2}$) \\
\hline
\multicolumn{5}{l}{{\bf 2012}} \\
13850 &SH&  60.0 &  59.28 & 2012-02-06 00:37:28 & 1 - 0 \\
14392 &SH&  60.0 &  58.47 & 2012-02-09 06:17:03 & 5 - 1 & 2 & 55966.6022 & 55966.6167 &  1250 & $0.041\pm0.014$ & 46.6 \\ 
      & &       &        &                     &       & 3 & 55966.6167 & 55966.6553 &  3336 & $0.102\pm0.018$ & 396.0 \\
      & &       &        &                     &       & 4 & 55966.6553 & 55966.6669 &   997 & $0.031\pm0.009$ & 27.0 & {\bf 469.4} \\
14394 &SH&  18.0 &  17.83 & 2012-02-10 03:16:18 & 1 - 0 \\                                                          
14393 &SH&  42.0 &  41.0  & 2012-02-11 10:13:03 & 1 - 0 \\                                                          
13856 &SH&  40.0 &  39.54 & 2012-03-15 08:45:22 & 1 - 0 \\                                                          
13857 &SH&  40.0 &  39.04 & 2012-03-17 08:57:45 & 1 - 0 \\                                                          
13854 &SH&  25.0 &  22.76 & 2012-03-20 10:12:13 & 9 - 4 & 2 & 56006.4881 & 56006.4943 &   538 & $0.039\pm0.014$ & 18.7 & {\bf 18.7} \\
      & &       &        &                     &       & 4 & 56006.527  & 56006.537  &   861 & $0.028\pm0.011$ & 19.8 & {\bf 19.8} \\
      & &       &        &                     &       & 6 & 56006.5853 & 56006.5963 &   949 & $0.024\pm0.008$ & 18.8 & {\bf 18.8} \\
      & &       &        &                     &       & 8 & 56006.6803 & 56006.688  &   669 & $0.042\pm0.016$ & 25.9 & {\bf 25.9} \\
14413 &SH&  15.0 &  14.53 & 2012-03-21 06:44:50 & 1 - 0 \\                                                          
13855 &SH&  20.0 &  19.8  & 2012-03-22 11:24:50 & 1 - 0 \\                                                          
14414 &SH&  20.0 &  19.8  & 2012-03-23 17:48:38 & 1 - 0 \\                                                          
13847 &SH& 157.0 & 152.05 & 2012-04-30 16:16:52 & 3 - 1 & 2 & 56048.5111 & 56048.5471 &  3111 & $0.013\pm0.003$ & 28.4 & {\bf 28.4} \\
14427 &SH&  80.0 &  79.01 & 2012-05-06 20:01:01 & 5 - 2 & 2 & 56054.1274 & 56054.1479 &  1776 & $0.015\pm0.006$ & 20.8 & {\bf 20.8} \\
      & &       &        &                     &       & 4 & 56054.4684 & 56054.5067 &  3309 & $0.009\pm0.003$ & 19.3 & {\bf 19.3} \\
13848 &SH& 100.0 &  96.87 & 2012-05-09 12:02:49 & 1 - 0 \\                                                          
13849 &SH& 180.0 & 176.41 & 2012-05-11 03:18:41 & 7 - 3 & 2 & 56058.6909 & 56058.7049 &  1204 & $0.015\pm0.007$ & 13.6 & {\bf 13.6} \\
      & &       &        &                     &       & 4 & 56059.0214 & 56059.0469 &  2203 & $0.011\pm0.006$ & 16.7 & {\bf 16.7} \\
      & &       &        &                     &       & 6 & 56060.1334 & 56060.1681 &  3000 & $0.028\pm0.006$ & 70.0 & {\bf 70.0} \\
13846 &SH&  57.0 &  55.47 & 2012-05-16 10:41:16 & 1 - 0 \\                                                          
14438 &SH&  26.0 &  25.46 & 2012-05-18 04:28:40 & 1 - 0 \\                                                          
13845 &SH& 135.0 & 133.54 & 2012-05-19 10:42:32 & 3 - 1 & 2 & 56067.8636 & 56067.884  &  1761 & $0.036\pm0.010$ & 55.3 & {\bf 55.3} \\ 
14460 &SH&  24.0 &  23.75 & 2012-07-09 22:33:03 & 1 - 0 \\                                                          
13844 &SH&  20.0 &  19.8  & 2012-07-10 23:10:57 & 1 - 0 \\                                                          
14461 &SH&  51.0 &  50.3  & 2012-07-12 05:48:45 & 1 - 0 \\                                                          
13853 &SH&  74.0 &  72.71 & 2012-07-14 00:37:17 & 1 - 0 \\                                                          
13841 &SH&  45.0 &  44.48 & 2012-07-17 21:06:39 & 1 - 0 \\                                                          
14465 &SH&  44.0 &  43.77 & 2012-07-18 23:23:38 & 4 - 2 & 1 & 56126.9761 & 56127.0274 &  4433 & $0.0099\pm0.0017$ & 29.7 & {\bf 29.7} \\
      & &       &        &                     &       & 3 & 56127.1777 & 56127.2023 &  2130 & $0.009\pm0.005$ & 12.7 & {\bf 12.7} \\
14466 &SH&  45.0 &  44.49 & 2012-07-20 12:37:09 & 2 - 1 & 1 & 56128.5412 & 56128.5545 &  1146 & $0.024\pm0.005$ & 22.3 & {\bf 22.3} \\
13842 &SH& 192.0 & 189.25 & 2012-07-21 11:52:41 & 7 - 3 & 2 & 56130.1889 & 56130.2239 &  3026 & $0.022\pm0.005$ & 49.5 & {\bf 49.5} \\
      & &       &        &                     &       & 4 & 56130.9089 & 56130.9176 &   755 & $0.028\pm0.011$  & 18.1& {\bf 18.1} \\
      & &       &        &                     &       & 6 & 56131.4921 & 56131.5827 &  7830 & $0.0107\pm0.0019$ & 58.0 & {\bf 58.0} \\
13839 &SH& 180.0 & 173.95 & 2012-07-24 07:02:59 & 5 - 2 & 2 & 56132.3879 & 56132.3985 &   917 & $0.029\pm0.03$ & 23.5 & {\bf 23.5} \\
      & &       &        &                     &       & 4 & 56134.0027 & 56134.0374 &  2998 & $0.050\pm0.008$ & 143.7 & {\bf 143.7} \\
13840 &SH& 163.0 & 160.39 & 2012-07-26 20:01:52 & 3 - 1 & 2 & 56136.48   & 56136.5078 &  2399 & $0.010\pm0.005$ & 14.8 & {\bf 14.8} \\
14432 &SH&  75.0 &  73.3  & 2012-07-30 12:56:02 & 3 - 2 & 1 & 56138.557  & 56138.6163 &  5130 & $0.0070\pm0.0013$ & 22.5 & {\bf 22.5} \\ 
      & &       &        &                     &       & 3 & 56139.3726 & 56139.4166 &  3803 & $0.030\pm0.006$ & 99.6 & {\bf 99.6} \\ 
13838 &SH& 100.0 &  98.26 & 2012-08-01 17:29:25 & 4 - 1 & 2 & 56141.013  & 56141.0298 &  1452 & $0.05\pm0.02$ & 68.4 \\
      & &       &        &                     &       & 3 & 56141.0298 & 56141.0477 &  1544 & $0.012\pm0.006$ & 12.5 & {\bf 80.9}\\
13852 &SH& 155.0 & 154.52 & 2012-08-04 02:37:36 & 7 - 3 & 2 & 56143.3157 & 56143.3309 &  1312 & $0.029\pm0.007$ & 32.9 & {\bf 32.9} \\
      & &       &        &                     &       & 4 & 56143.7398 & 56143.8904 & 13008 & $0.0049\pm0.0015$ & 31.5 & {\bf 31.5} \\
      & &       &        &                     &       & 6 & 56144.3311 & 56144.3691 &  3282 & $0.009\pm0.004$ & 21.7 & {\bf 21.7} \\
14439 &SH& 112.0 & 110.27 & 2012-08-06 22:16:59 & 3 - 1 & 2 & 56147.1319 & 56147.1528 &  1807 & $0.012\pm0.005$ & 15.9 & {\bf 15.9} \\
14462 &SH& 134.0 & 131.64 & 2012-10-06 16:31:54 & 5 - 2 & 2 & 56207.1717 & 56207.2055 &  2924 & $0.008\pm0.005$ & 14.5 & {\bf 14.5}\\
      & &       &        &                     &       & 4 & 56208.187  & 56208.2136 &  2303 & $0.013\pm0.006$ & 22.3 & {\bf 22.3} \\
14463 &SH&  31.0 &  30.37 & 2012-10-16 00:52:28 & 3 - 1 & 2 & 56216.2406 & 56216.2466 &   522 & $0.069\pm0.019$ & 36.4 & {\bf 36.4} \\
13851 &SH& 107.0 & 105.66 & 2012-10-16 18:48:46 & 6 - 1 & 2 & 56217.8153 & 56217.8237 &   732 & $0.020\pm0.012$ & 12.2 \\
      & &       &        &                     &       & 3 & 56217.8237 & 56217.8406 &  1456 & $0.087\pm0.014$ & 136.8 \\
      & &       &        &                     &       & 4 & 56217.8406 & 56217.8582 &  1522 & $0.040\pm0.013$ & 55.4 \\
      & &       &        &                     &       & 5 & 56217.8582 & 56217.8759 &  1527 & $0.011\pm0.008$ & 11.4 & {\bf 215.8} \\
15568 &SH&  36.6 &  35.59 & 2012-10-18 08:55:23 & 2 - 1 & 2 & 56218.7394 & 56218.8039 &  5570 & $0.006\pm0.007$ & 18.6 & {\bf 18.6} \\
13843 &SH& 121.0 & 119.1  & 2012-10-22 16:00:48 & 5 - 1 & 2 & 56223.3833 & 56223.3929 &   833 & $0.013\pm0.015$ & 8.0 \\
      & &       &        &                     &       & 3 & 56223.3929 & 56223.4178 &  2153 & $0.053\pm0.007$ & 110.6  \\
      & &       &        &                     &       & 4 & 56223.4178 & 56223.4785 &  5241 & $0.008\pm0.003$ & 24.8 & {\bf 143.4} \\
15570 &SH&  69.4 &  67.81 & 2012-10-25 03:30:43 & 4 - 1 & 2 & 56225.2334 & 56225.255  &  1869 & $0.011\pm0.005$ & 15.1 \\
      & &       &        &                     &       & 3 & 56225.255  & 56225.2614 &   547 & $0.037\pm0.017$ & 18.2 & {\bf 33.3} \\
14468 &SH& 146.0 & 144.15 & 2012-10-29 23:42:08 & 5 - 2 & 2 & 56230.3007 & 56230.3858 &  7350 & $0.008\pm0.004$ & 40.0 & {\bf 40.0} \\
      & &       &        &                     &       & 4 & 56231.568  & 56231.5908 &  1971 & $0.016\pm0.006$ & 23.8 & {\bf 23.8} \\
\hline                                                                   
\end{tabular}                                                            
\caption{List of all \chandra\ observations on \sgras\ performed in 2012. 
Same as previous table. }
\label{ExpC2}
\end{center}
\end{table*} 

\begin{table*}
\scriptsize
\begin{center}
\begin{tabular}{ | c c r r c c l l l r c r r }
\hline
OBSID &Obs& Exp & Clean        & Obs            & Nb - Nf & Nb & Block & Block & Dur & Observed & Block          & Flare \\
            &mode&       &  Exp           & date            &             &       &Start  & Stop  &        & Count rate & Fluence     & Fluence \\
            &         &       &                  &                    &             &       &          &          &       &                    & 10$^{-10}$          & 10$^{-10}$ \\
            &         &(ks) & (ks)           &                    &             &       &(MJD)&(MJD) & (s) &(ph~s$^{-1}$)& (erg~cm$^{-2}$) & (erg~cm$^{-2}$) \\
\hline
\multicolumn{5}{l}{{\bf 2013}} \\
14941 &I&  20.0 &  19.82 & 2013-04-06 01:22:20 & 1 - 0 \\
14942 &I&  20.0 &  19.83 & 2013-04-14 15:41:48 & 1 - 0 \\
14702 &S& 15.0 &  13.67 & 2013-05-12 10:37:43 & 1 - 0 \\
15040 &SH&25.0&  23.75 & 2013-05-25 11:37:30 & 1 - 0 \\
14703 &S& 20.0 &  16.84 & 2013-06-04 08:44:10 & 1 - 0 \\
15651 &SH&15.0& 13.76 & 2013-06-05 21:31:31 & 1 - 0\\
15654 &SH&10.0&   9.03 & 2013-06-09 04:25:10 & 1 - 0 \\
14946 &S&  20.0 &  18.2  & 2013-07-02 06:48:37 & 1 - 0 \\
15041 &S&  50.0 &  45.41 & 2013-07-27 01:26:11 & 3 - 1 & 2 & 56500.1466 & 56500.1581 & 996 & $0.037\pm0.013$ & 6.3 & {\bf 6.3} \\
15042 &S&  50.0 &  45.67 & 2013-08-11 22:56:52 & 3 - 1 & 2 & 56516.3568 & 56516.484  & 10991 & $0.014\pm0.007$ & 14.9 & {\bf 15.1} \\
14945 &S&  20.0 &  18.2  & 2013-08-31 10:11:39  & 2 - 1 & 2 & 56535.6685 & 56535.6771 & 742 & $0.0256\pm0.009$ & 2.9 & {\bf 2.9}\\
15043 &S&  50.0 &  45.41 & 2013-09-14 00:03:46 & 9 - 1 & 2 & 56549.0847 & 56549.0898 & 447 & $0.17\pm0.04$ & 22.0 \\
      &     &  &        &                     &       & 3 & 56549.0898 & 56549.0938 & 341  & $0.48\pm0.11$ & 59.5 \\
      &     &  &        &                     &       & 4 & 56549.0938 & 56549.1069 & 1129 & $0.87\pm0.07$ & 428.9 \\
      &     &  &        &                     &       & 5 & 56549.1069 & 56549.1096 & 235 & $0.43\pm0.19$ & 35.9 \\
      &     &  &        &                     &       & 6 & 56549.1096 & 56549.123  & 1160 & $0.82\pm0.04$ & 401.7 \\
      &     &  &        &                     &       & 7 & 56549.123  & 56549.1333 & 886 & $0.39\pm0.03$ & 120.3 \\
      &     &  &        &                     &       & 8 & 56549.1333 & 56549.1488 & 1339  & $0.12\pm0.02$ & 43.0 & {\bf 1111.3} \\
14944 &S&  20.0 &  18.2  & 2013-09-20 07:01:50 & 3 - 1 & 2 & 56555.4619 & 56555.4855 & 2032 & $0.022\pm0.008$ & 6.7 & {\bf 6.7}\\
15044 &S&  50.0 &  42.69 & 2013-10-04 17:23:41 & 1 - 0 \\	
14943 &S&  20.0 &  18.2  & 2013-10-17 15:39:58 & 2 - 0.5 & 1 & 56582.6692 & 56582.6832 & 1212 & $0.022\pm0.008$ & 3.7 & {\bf 3.7} \\
14704 &S&  40.0 &  36.34 & 2013-10-23 08:53:23 & 1 - 0 \\
15045 &S&  50.0 &  45.41 & 2013-10-28 14:30:07 & 5 - 2 & 2 & 56593.675  & 56593.7017 & 2300 & $0.033\pm0.008$ & 12.3 & {\bf 12.3} \\
      &    &  &        &                     &       & 4 & 56593.831  & 56593.842  & 947 & $0.031\pm0.015$ & 4.8 & {\bf 4.8} \\
\multicolumn{5}{l}{{\bf 2014}} \\
16508 &S&  50.0 &  43.41 & 2014-02-21 11:36:41 & 2 - 0.5 & 2 & 56710.0238 & 56710.049  & 2173 & $0.024\pm0.006$ & 8.6 & {\bf 8.6}\\
16211 &S&  50.0 &  41.78 & 2014-03-14 10:17:20 & 1 - 0 \\
16212 &S&  50.0 &  45.41 & 2014-04-04 02:25:20 & 1 - 0 \\
16213 &S&  50.0 &  44.96 & 2014-04-28 02:43:58 & 1 - 0 \\
16214 &S&  50.0 &  45.41 & 2014-05-20 00:18:05 & 1 - 0 \\
16210 &S&  20.0 &  17.02 & 2014-06-03 02:58:16 & 1 - 0 \\
16597 &S&  20.0 &  16.46 & 2014-07-04 20:47:06 & 1 - 0 \\
16215 &S&  50.0 &  41.45 & 2014-07-16 22:42:45 & 1 - 0 \\
16216 &S&  50.0 &  42.69 & 2014-08-02 03:30:34 & 1 - 0 \\
16217 &S&  40.0 &  34.53 & 2014-08-30 04:49:05 & 3 - 1 & 2 & 56899.4921 & 56899.546  & 4661 & $0.020\pm0.004$ & 13.6 & {\bf 13.6}\\
16218 &S&  40.0 &  36.34 & 2014-10-20 08:21:21 & 9 - 2 & 2 & 56950.555  & 56950.5683 & 1145 & $0.062\pm0.014$ & 17.0 \\
      &     &  &        &                     &       & 3 & 56950.5683 & 56950.5725 & 369 & $0.17\pm0.08$ & 18.8 \\
      &     &  &        &                     &       & 4 & 56950.5725 & 56950.5822 & 831 & $0.46\pm0.07$ & 139.7 \\
      &     &  &        &                     &       & 5 & 56950.5822 & 56950.5898 & 660 & $0.27\pm0.07$ & 57.1 \\
      &     &  &        &                     &       & 6 & 56950.5898 & 56950.5935 & 314 & $0.10\pm0.05$ & 7.2 & {\bf 239.8} \\
      &     &  &        &                     &       & 8 & 56950.6187 & 56950.6295 & 933  & $0.046\pm0.016$ & 8.4 & {\bf 8.4} \\
\hline
\end{tabular}
\caption{List of all \chandra\ observations on \sgras\ performed in 2013 and 2014. Same as previous tables. }
\label{ExpC}
\end{center}
\end{table*} 

\begin{table*}
\scriptsize
\begin{center}
\begin{tabular}{ | c c c c c c c c c c c c c c c c c }
\hline
OBSID           & Obs Date                    & t$_{in}$ & t$_{fin}$ & Exp & Nb - Nf & Nb & Bstart & Bstop & Dur & Block         & Block Fluence & Flare Fluence\\
                      &                                    &              &               &        &              &         &          &         &         & count rate   & 10$^{-10}$          & 10$^{-10}$ \\
                      &                                   &(ks)        &(ks)         &(ks)  &               &       &(MJD)&(MJD) & (s) &(ph~s$^{-1}$)& (erg~cm$^{-2}$) & (erg~cm$^{-2}$) \\
\hline
  0112970601 &  2000-09-17 17:46:18  &0 &27.8   & 27.8  & 1 - 0 & \\
  0112972101 &  2001-09-04 01:20:42  &0 &26.7   & 26.7  & 2 - 1 &2 & 52156.3554 & 52156.3607 &  462 & $0.18\pm0.05$ & 10.4 & {\bf 10.4} \\
  0111350101 &  2002-02-26 03:16:43  &0 &52.8   & 52.8  & 1 - 0 & \\                                                          
  0111350301 &  2002-10-03 06:54:11  &0 &17.3   & 17.3  & 7 - 1 &2 & 52550.4244 & 52550.4287 &  372 & $0.26\pm0.05$ & 18.7 \\
             &                      &    &  &             &       &3 & 52550.4287 & 52550.4314 &  235 & $0.49\pm0.17$ & 27.8 \\ 
             &                      &    &  &             &       &4 & 52550.4314 & 52550.4427 &  969 & $0.80\pm0.06$ & 205.5 \\
             &                      &    &  &             &       &5 & 52550.4427 & 52550.4482 &  480 & $0.56\pm0.11$ & 64.1 \\ 
             &                      &    &  &             &       &6 & 52550.4482 & 52550.4535 &  452 & $0.27\pm0.08$ & 21.3 & {\bf 337.4} \\ 
  0202670501 &  2004-03-28 15:03:52  &13& 81.5  & 68.5  & 1 - 0 & \\                                                          
  0202670601 &  2004-03-30 14:46:36  &14.5& 77  & 62.5  & 5 - 0\dag & \\                                                          
  0202670701 &  2004-08-31 03:12:01  &42& 123   & 81    & 3 - 0\dag & \\                                                          
  0202670801 &  2004-09-02 03:01:39  &0 &125    &125    & 1 - 0 & \\                                                          
  0302884001 &  2006-09-08 16:56:48  &0 &6.9    &  6.9  & 1 - 0 & \\                                                          
  0302882601 &  2006-02-27 04:04:34  &0 &6.9    &  6.9  & 1 - 0 & \\                                                          
  0402430701 &  2007-03-30 21:05:17  &0 &34.2   & 34.2  & 1 - 0 & \\                                                          
  0402430301 &  2007-04-01 14:45:02  &3.5&90.5  & 87    & 2 - 0 & \\                                                          
  0402430401 &  2007-04-03 14:32:24  &0 &82     & 82    &10 - 3 &2 & 54194.2202 & 54194.2314 &  964 & $0.20\pm0.05$ & 24.3 \\
             &                      &    &  &             &       &3 & 54194.2314 & 54194.2486 & 1490 & $0.53\pm0.03$ & 185.6 \\ 
             &                      &    &  &             &       &4 & 54194.2486 & 54194.258  &  808 & $0.28\pm0.05$ & 41.2 & {\bf 251.1} \\
             &                      &    &  &             &       &6 & 54194.4816 & 54194.4836 &  175 & $0.29\pm0.11$ & 9.3 & {\bf 9.3} \\
             &                      &    &  &             &       &8 & 54194.6041 & 54194.6147 &  914 & $0.17\pm0.05$ & 16.8 \\
             &                      &    &  &             &       &9 & 54194.6147 & 54194.6245 &  845 & $0.26\pm0.05$ & 40.1 \\
             &                      &    &  &             &      &10& 54194.6245 & 54194.6439 &1674 & $0.15\pm0.04$ & 10.6 & {\bf 67.5} \\
  0504940201 &  2007-09-06 10:05:50  &0 &13     & 13    & 1 - 0 & \\                                                          
  0511000301 &  2008-03-03 23:25:56  &0 &6.9    &  6.9  & 1 - 0 & \\                                                          
  0505670101 &  2008-03-23 14:59:43  &0 &105.7  &105.7  & 1 - 0 & \\                                                          
  0511000401 &  2008-09-23 15:15:50  &0 &6.9    &  6.9  & 1 - 0 & \\                                                          
  0554750401 &  2009-04-01 00:55:25  &0 &39.9   & 39.9  & 1 - 0 & \\                                                          
  0554750501 &  2009-04-03 01:33:27  &0 &44.3   & 44.3  & 1 - 0 & \\                                                          
  0554750601 &  2009-04-05 02:17:13  &0 &39.1   & 39.1  & 1 - 0 & \\                                                          
  0604300601 &  2011-03-28 07:49:58  &0 &34     & 34    & 1 - 0 & \\                                                          
  0604300701 &  2011-03-30 07:44:39  &0 &48.9   & 48.9  & 3 - 1 &2 & 55650.7392 & 55650.7622 & 1990 & $0.215\pm0.014$ & 70.2 & {\bf 70.2} \\
  0604300801 &  2011-04-01 07:48:13  &0 &34.5   & 34.5  & 1 - 0 & \\                                                          
  0604300901 &  2011-04-03 07:52:07  &0 &24.5   & 24.5  & 2 - 1 &1 & 55654.3445 & 55654.3637 & 1662 & $0.183\pm0.015$ & 42.0 & {\bf 42.0} \\
  0604301001 &  2011-04-05 07:09:33  &0 &45     & 45    & 1 - 0 & \\                                                          
  0658600101 &  2011-08-31 23:14:23  &0 &49.9   & 49.9  & 1 - 0 & \\                                                          
  0658600201 &  2011-09-01 20:03:48  &0 &41     & 41    & 1 - 0 & \\                                                          
  0674600601 &  2012-03-13 03:52:36  &0 &21.5   & 21.5  & 1 - 0 & \\                                                          
  0674600701 &  2012-03-15 04:47:06  &0 &15.9   & 15.9  & 1 - 0 & \\                                                          
  0674601101 &  2012-03-17 02:30:16  &7.2&18.3  & 11.1  & 1 - 0 & \\                                                          
  0674600801 &  2012-03-19 03:52:38  &0 &22.9   & 22.9  & 1 - 0 & \\                                                          
  0674601001 &  2012-03-21 03:30:40  &0 &23.9   & 23.9  & 1 - 0 & \\                                                          
  0694640301 &  2012-08-31 11:20:26  &0 &41.9   & 41.9  & 1 - 0 & \\                                                          
  0694641101 &  2012-09-24 10:16:44  &0 &41.9   & 41.9  & 1 - 0 & \\                                                          
  0743630201 &  2014-08-30 19:20:01  &0 &33.9   & 33.9  & 6 - 2 &2 & 56899.9876 & 56899.9982 &  914 & $0.34\pm0.05$ & 44.4 \\
             &                      &    &  &             &       &3 & 56899.9982 & 56900.0136 & 1334 & $0.71\pm0.07$ & 213.5 \\
             &                      &    &  &             &       &4 & 56900.0136 & 56900.0185 &  418 & $0.43\pm0.11$ & 31.5 & {\bf 289.4} \\
             &                      &    &  &             &       &6 & 56900.1886 & 56900.2056 & 1468 & $0.32\pm0.05$ & 62.5 & {\bf 62.5} \\
  0743630301 &  2014-08-31 20:23:30  &0 &26.9   & 26.9  & 4 - 1 &2 & 56901.0303 & 56901.0639 & 2905 & $0.23\pm0.02$ & 35.1 \\
             &                      &    &  &             &       &3 & 56901.0639 & 56901.0807 & 1454 & $0.34\pm0.03$ & 67.3 & {\bf 102.4} \\
  0743630401 &  2014-09-27 17:30:23  &0 &33.5   & 33.5  & 1 - 0 & \\                                                          
  0743630501 &  2014-09-28 21:01:46  &0 &39.2   & 39.2  & 3 - 1 &2 & 56929.2548 & 56929.2596 &  421 & $0.33\pm0.09$ & 19.1 \\
             &                      &    &  &             &       &3 & 56929.2596 & 56929.3434 & 7233 & $0.20\pm0.03$ & 39.7 & {\bf 58.8} \\
\hline
\end{tabular}
\caption{List of all \xmm\ observations with \sgras\ in the field of view, considered in this work (see \S~2). 
For each observation the observation ID, the observation start date, the start and end time, 
after filtering of bright background flares, the cleaned exposure, the number of different 
blocks and the number of the different flares detected, are reported. 
In the following columns we report information about each flaring block. 
We report the block number, the block start and stop times (in MJD), 
the duration (in seconds), the mean flux (in units of $10^{-13}$~erg~cm$^{-2}$~s$^{-1}$, 
with associated error, computed through bootstrap simulations) and the fluence (in units 
of $10^{-10}$~erg~cm$^{-2}$, after subtraction of the average contribution of the non 
flaring blocks) for each flaring block. The last column shows, for each flare, the total flare 
fluence in erg~cm$^{-2}$. \dag Eclipses of CXOGC~J174540.0-290031 are detected 
during these observations. 
}
\label{ExpXMM}
\end{center}
\end{table*} 

\section*{Acknowledgments}

The authors wish to thank Jan-Uwe Ness, Ignacio de la Calle and the rest of the \xmm\ 
scheduling team for the enormous support that made the new \xmm\ observations possible.
The authors also wish to thank the anonymous referee for the very useful comments 
that significantly improved the paper. 
We thank all the members of the \sgras's \chandra\ XVP collaboration (www.sgra-star.com) 
for contributing in collecting X-ray data and for compiling a legacy dataset for astronomy. 
GP thanks Alessandro Ballone for useful discussion. 
This research has made use both of data obtained with \xmm, an ESA 
science mission with instruments and contributions directly funded by ESA Member 
States and NASA, and on data obtained from the Chandra Data Archive.
GP acknowledges support via an EU Marie Curie Intra-European Fellowship under 
contract no. FP7-PEOPLE-2012-IEF-331095. 
The GC \xmm\ monitoring project is supported by the Bundesministerium 
f\"{u}r Wirtschaft und Technologie/Deutsches Zentrum f\"{u}r Luft- und Raumfahrt 
(BMWI/DLR, FKZ 50 OR 1408) and the Max Planck Society. 
DH acknowledges support from Chandra X-ray Observatory (CXO) 
Award Numbers GO3-14121X, G04-15091A, and GO4-15091C, 
operated by the Smithsonian Astrophysical Observatory for and 
on behalf of NASA under contract NAS8-03060, and also from NASA 
Swift grant NNX14AC30G. ND acknowledges support via an EU Marie 
Curie Intra-European fellowship under contract no. FP-PEOPLE-2013-IEF-627148. 
MC, AG, RT, SS acknowledge support from CNES.

\end{document}